\documentclass[pdflatex,iicol,sn-apa]{sn-jnl}
\usepackage[T1]{fontenc}
\usepackage{amsmath,amssymb,booktabs,tabularx,array,graphicx,xurl}
\usepackage{lmodern,microtype}
\graphicspath{{figures/}}
\makeatletter
\def\toclevel@paragraph{3}
\makeatother
\hypersetup{pdftitle={HP-GSBS: Accelerating greedy state boundary search with reusable interval statistics},pdfauthor={Yanzhe Zhao and Yongzhang Ji},colorlinks=true,allcolors=blue}
\newcommand{\mainfigure}[4][!tp]{\begin{figure*}[#1]\centering\includegraphics[width=\textwidth,height=.74\textheight,keepaspectratio]{#2}\caption{#3}\label{#4}\end{figure*}}
\newcommand{\figOneCaption}{Neural-state segmentation and the reusable statistics underlying HP-GSBS. (a) GSBS represents contiguous periods of neural activity by their mean spatial patterns, searches for state boundaries, and selects the number of states using the t-distance curve. (b) HP-GSBS uses temporal prefix sums to calculate the matching score of each interval, $Q_{a,b}=F(a,b)$. Adding one or two boundaries replaces the score of the original interval with the scores of the resulting intervals, allowing all other contributions to be reused. The best proposals are compared using t-distance, followed by sequential boundary refinement. (c) Within-state and adjacent-state correlations occupy identifiable regions of the time-by-time correlation matrix. Two-dimensional prefix tables of eligible-pair counts, correlations, and squared correlations allow their summary statistics to be retrieved using four-corner rectangle queries. The full symmetric matrix is shown for illustration; each eligible pair contributes only once to the statistics. Three numerical backends supply reusable statistics to the same sequential search controller. Data-derived elements use a 120-TR excerpt of the Sherlock fMRI dataset \citep{baldassano2020sherlock,chen2017shared}; heatmaps display feature subsets, while all 4,230 voxels are used for fitting. Construction details and data provenance are given in Online Resource~1, Section~S4.}

\newcommand{\figTwoCaption}{Segmentation of a simulated recording with known state boundaries. (a) The input contains four states; dashed lines indicate the three generating boundaries. (b) The official GSBS implementation and all three HP-GSBS backends identify the same boundaries at the selected state count ($K=4$). (c) Complete t-distance curves, retaining the reference double-insertion indexing; all implementations select four states. (d) Absolute differences between HP-GSBS and the official t-distance values; the two GPU curves overlap. Simulation parameters: seed 19, Gaussian noise standard deviation 0.05, $T=80$, $P=16$, and $K_{\max}=12$.}

\newcommand{\figThreeCaption}{Agreement between implementations for a complete 64-channel EEG recording. (a) Channel-standardized EEG and the selected sequence of 126 states; alternating colors distinguish neighboring intervals. (b) An enlarged window (120.51--140.43 s) showing the activity patterns and the identical boundary locations identified by the four implementations. Both EEG heatmaps use a color scale clipped at $\pm3$. (c) Spatial-pattern correlations between time points in the same window, shown as a full symmetric matrix; outlined blocks indicate the selected state intervals. (d) Complete evaluated t-distance curves, retaining the reference double-insertion indexing; markers are staggered for visibility. All implementations select 126 states with identical complete boundary histories (maximum $|\Delta\mathrm{TD}|=2.43\times10^{-13}$). The recording is sampled at 5.12 Hz; settings are $K_{\max}=131$, $d_{\min}=5$, statewise detection, sequential refinement, block size 50, and FP64.}

\newcommand{\figFourCaption}{Complete-fit runtime and speedup across controlled benchmark configurations. (a--c) The recording length $T$, feature count $P$, and maximum state count $K_{\max}$ are varied separately, holding the other two factors fixed. Filled points show the two input-level observations, open markers their median, and whiskers their observed range. Each HP-GSBS observation is the median of three timed fits after warm-up; each official observation is a single timed fit. Panel b annotates the runtimes for the central configuration. (d--f) Speedup ratios compare the official and optimized runtimes on the same input and are summarized across the two seeds. Panel d highlights the median PyTorch speedup at $T=768$. The central configuration is shared across panels. All fits use statewise detection and sequential refinement.}

\newcommand{\figFiveCaption}{Effects of search optimization, memory-efficient computation, and recording-level parallelism. (a) Stepwise CPU ablation on a $1{,}336\times64$ EEG input ($K_{\max}=131$), starting from a configuration that precomputes interval scores and stores all candidate scores (Materialized). (b) Left: peak process resident set size (RSS) and complete-fit time for Full-Q and low-memory execution on a $5{,}000\times8$ input. Right: peak RSS across four low-memory workloads in which both recording length and feature dimension increase. The $5{,}000\times8$ low-memory input has peak RSS values of 190.6 and 191.1 MiB in the separate experiments shown on the left and right, respectively. All measurements in b use single-boundary search, no refinement, block size zero, and $K_{\max}=16$. (c,d) Completed fits per minute for CPU workers and GPU batches on short synthetic inputs and a repeated EEG input requiring a larger state-count search. CPU batches contain 32 and eight tasks, respectively. Runtimes in a and b and throughputs in c and d are three-run medians with observed-range whiskers; memory values are measured peaks. Complete settings and numerical results are given in Online Resource~1 (Tables~S9--S11) and the accompanying data archive.}

\newcommand{\figSixCaption}{Neural-state segmentation and alignment of estimated boundaries across viewers of the same movie. (a) Group-mean angular-gyrus activity and selected group boundaries. The heatmap displays 72 evenly sampled voxels, spatially z-scored at each time point with colors clipped at $\pm3$; fitting uses all 4,230 voxels. (b) Selected boundary locations for the group response and 16 individual participants on a common timeline; participant numbering follows the supplied array. (c) Complete evaluated t-distance curves and the selected state-count indices. The vertical line marks $K_{\max}=60$; the reference double-insertion convention permits index 61. (d) Distribution of cohort-mean pairwise $F_1$ after independently circularly shifting each participant's boundary sequence, with the observed $F_1$ marked. One-to-one matching uses a tolerance of $\pm3$ TR ($\pm4.5$ s); 1,000 null draws yield $p=0.001$. The data cover 23.65 min of the movie at TR = 1.5 s.}

\begin{document}
\title[HP-GSBS: Reusable interval statistics]{HP-GSBS: Accelerating greedy state boundary search with reusable interval statistics}
\author*[1]{\fnm{Yanzhe} \sur{Zhao}}\email{justinzhaoyz@outlook.com}
\author[2]{\fnm{Yongzhang} \sur{Ji}}
\affil[1]{\orgdiv{School of Future Technology}, \orgname{Tianjin University}, \orgaddress{\city{Tianjin}, \country{China}}}
\affil[2]{\orgdiv{Faculty of Psychology}, \orgname{Southwest University}, \orgaddress{\city{Chongqing}, \country{China}}}
\abstract{Identifying transitions between relatively stable patterns of neural activity helps reveal how the brain organizes continuous experience. Greedy State Boundary Search (GSBS) identifies such transitions directly from neural recordings, but repeatedly reconstructing state templates and evaluating correlations between time points makes large-scale analyses computationally expensive. We introduce HP-GSBS, an open-source implementation based on an exact mathematical reformulation. We express the template-matching objective as a sum of interval scores obtained from temporal prefix sums. We also compute Welch t-distance (TD), the criterion used to compare segmentations, from two-dimensional cumulative correlation tables using four lookups per rectangular region. Efficient candidate search, local refinement, and caching further reduce repeated work across CPU, PyTorch CUDA, and Triton backends. Across controlled benchmarks varying recording length, feature dimension, and search depth, HP-GSBS achieved 154- to 1,730-fold speedups while reproducing the reference boundaries at every evaluated stage and selecting identical state counts (maximum $|\Delta\mathrm{TD}|=1.88\times10^{-11}$). The three backends also agreed on four electroencephalography (EEG) recordings from the Brain, Body, and Behavior Dataset (120--173 selected states). In an extended benchmark, a 1.7-hour reference EEG fit took 0.84--1.61 s with HP-GSBS. For movie-viewing functional magnetic resonance imaging (fMRI) data from 16 participants, group and individual segmentations of 4,230 angular-gyrus voxels took 213 s in total on a single CPU. Boundaries aligned across viewers more closely than expected under a circular-shift null model (mean pairwise $F_1=0.414$ versus 0.325, $p=0.001$). Parallel analysis of independent recordings and an optional low-memory mode facilitate large-scale, multi-timescale neural-state studies on standard laboratory hardware.}

\maketitle
\noindent\small Yanzhe Zhao: \href{https://orcid.org/0009-0003-5342-6611}{ORCID 0009-0003-5342-6611}\normalsize
\section{Introduction}\label{sec:introduction}
Continuous experience unfolds at multiple timescales. During a film, a conversation, or a familiar activity, sensory information changes from moment to moment, while the context of an ongoing event may remain relatively stable. Multivariate neural recordings provide a way to study this organization by tracking the pattern of activity across recorded features, such as functional magnetic resonance imaging (fMRI) voxels or electroencephalography (EEG) channels. We refer to a contiguous time interval with a relatively stable spatial activity pattern as a neural state, and to the transition between two such intervals as a state boundary. Naturalistic fMRI studies have related neural state boundaries to narrative structure and memory \citep{baldassano2017}, motivating methods that estimate neural event structure directly from recorded brain activity. EEG provides a complementary way to examine temporal organization during continuous experience, including educational video viewing \citep{bbbd2026}. The resulting state sequence describes when transitions occur, how long states last, and which activity patterns characterize them. Since a naturalistic stimulus contains both rapid perceptual changes and slower changes in narrative context, meaningful temporal structure may emerge at several levels of granularity. Examining how estimated boundaries change as the number of states increases can therefore help characterize temporal organization across brain regions, stimuli, and individuals.

Estimating neural states from a continuous recording involves two related questions: where the transitions occur and how many states best describe the recording. Greedy State Boundary Search (GSBS) addresses the first question by iteratively dividing a time-by-feature matrix into states \citep{geerligs2021}. For each candidate boundary, the algorithm measures how closely the observed activity patterns resemble the mean patterns of the states to which they are assigned. GSBS addresses the second question using a criterion called t-distance, which compares the similarity of activity patterns within a state with their similarity across neighboring states. A subsequent extension, known as statewise detection, allows the algorithm to propose two boundaries within an existing state in a single iteration. This is particularly useful when a short-lived activity pattern occurs between longer periods of activity, because the onset and offset of the new state can be identified together \citep{geerligs2022}. GSBS has been used to characterize cortical temporal organization, including a partially nested hierarchy of neural states across brain regions. By retaining the segmentation and t-distance at each evaluated state count, researchers can compare temporal structure at different levels of granularity as well as examine shared and individual patterns of state boundaries.

The computational cost of GSBS, however, limits the scale at which these analyses can be performed. In the original implementation, a fit to a long recording may take tens of minutes or more than an hour; for example, fitting one EEG recording with 1,336 time points took approximately 1.7 hours (Online Resource~1, Table~S8). Repeating such fits across brain regions, participants, and parameter settings makes whole-brain analyses, sensitivity analyses, and permutation-based procedures that require refitting difficult to conduct on standard hardware.

Much of this cost arises because the algorithm repeatedly calculates quantities that remain unchanged during the search. Consider a recording that has already been divided into several states. When GSBS evaluates a new boundary within one state, only that state is split; the observations and mean patterns of the other states remain the same. Yet the reference-style calculation repeatedly reconstructs templates and evaluates correlations for candidate partitions. Searching for two boundaries at once further increases this cost because the number of possible pairs grows quadratically with the length of the searched interval. A similar source of redundancy occurs when t-distance is evaluated after a partition update: many of the same correlations between time points are collected again even though the states to which those time points belong have not changed.

These observations suggest a way to accelerate GSBS: separate the quantities that depend only on the recorded data from those that change as boundaries are added or moved. For template matching, we show that the overall score can be expressed as a sum of contributions from individual state intervals. Each contribution can be calculated from temporal prefix sums, which store cumulative activity statistics. Inserting a boundary then requires replacing only the contribution of the state being split with those of the resulting states. For t-distance, we exploit the fact that the correlation distributions can be summarized by their counts, sums, and squared sums. Since each state occupies a contiguous time interval, these summaries can be retrieved from precomputed tables of correlations between time points. Both reformulations preserve the mathematical criteria used by GSBS.

We combine these reusable statistics with a search procedure that retains the best candidate as candidates are evaluated, refines boundaries within the specified local window, and reuses search results from unchanged groups of consecutive states. The separation between constructing the statistics and making sequential boundary decisions also allows CPU and GPU backends to share the same search procedure.

Here, we introduce HP-GSBS, an open-source implementation of these reformulations \citep{hpgsbs2026}. We first derive the reusable interval and correlation statistics and explain how they are incorporated into the GSBS search. Using controlled synthetic benchmarks, we then examine whether the optimized implementations reproduce the original segmentation results and how their runtime varies with recording length, feature dimension, and search depth. Next, we examine numerical agreement between the CPU implementation and the two GPU backends, PyTorch CUDA and Triton, using four multichannel EEG recordings from the Brain, Body, and Behavior Dataset (BBBD; \citealp{bbbd2026}). These recordings require searches over more than one hundred states. Finally, we apply HP-GSBS to naturalistic movie-viewing fMRI data from 16 participants, each with 4,230 angular-gyrus voxels, to illustrate how faster individual-level segmentation can support investigations of shared neural state boundaries across viewers.

\section{Methods}\label{sec:methods}
\subsection{GSBS formulation and computational structure}
The input to GSBS is a matrix $X\in\mathbb{R}^{T\times P}$, where each row $x_t$ represents the spatial activity pattern at time point $t$. The $P$ features may correspond, for example, to fMRI voxels or EEG channels. GSBS divides the recording into consecutive time intervals, each representing a neural state. We denote a segmentation by $\mathcal S$ and represent each state as a half-open interval $[a,b)$ with mean activity pattern
\[
m_{ab}=(b-a)^{-1}\sum_{t=a}^{b-1}x_t.
\]
This mean pattern serves as the template for the state. A segmentation should assign observations to states whose templates closely resemble the activity patterns observed at those time points. GSBS therefore evaluates the mean spatial Pearson correlation between each observation and its assigned state template:
\begin{equation}
J(\mathcal S)=\frac{1}{T}\sum_{[a,b)\in\mathcal S}\sum_{t=a}^{b-1}\operatorname{corr}(x_t,m_{ab}).
\label{eq:objective}
\end{equation}
Thus, $J(\mathcal S)$ quantifies how well the state templates represent the observed spatial patterns throughout the recording.

GSBS adds boundaries iteratively. At each step, it searches for a single new boundary or, when statewise detection is enabled, also considers a pair of new boundaries within an existing state. To organize the search, GSBS can group consecutive states into local blocks. It first identifies the best candidates within each block using scores normalized by the block length and then compares those candidates over the full recording using Eq.~\eqref{eq:objective}. When both single- and double-boundary proposals are available, GSBS compares their t-distances to choose which proposal to accept. When refinement is enabled, existing boundaries are then reconsidered sequentially. The t-distance values recorded throughout the search form a curve over the evaluated state counts, whose maximum determines the selected number of states \citep{geerligs2021,geerligs2022}.

Two different calculations therefore recur during the search. The first involves comparing observations with state templates to identify promising boundary locations. The second involves comparing correlations between time points to evaluate a segmentation. Both calculations repeatedly access information that remains unchanged during a local boundary update. In Section~\ref{sec:intervals}, we show how to reuse the template-matching scores of individual intervals. In Section~\ref{sec:td}, we show how to obtain t-distance from precomputed summaries of pairwise correlations (Fig.~\ref{fig:architecture}). These reformulations retain the reference candidate domain, proposal ordering, and sequential partition updates.

\mainfigure[!t]{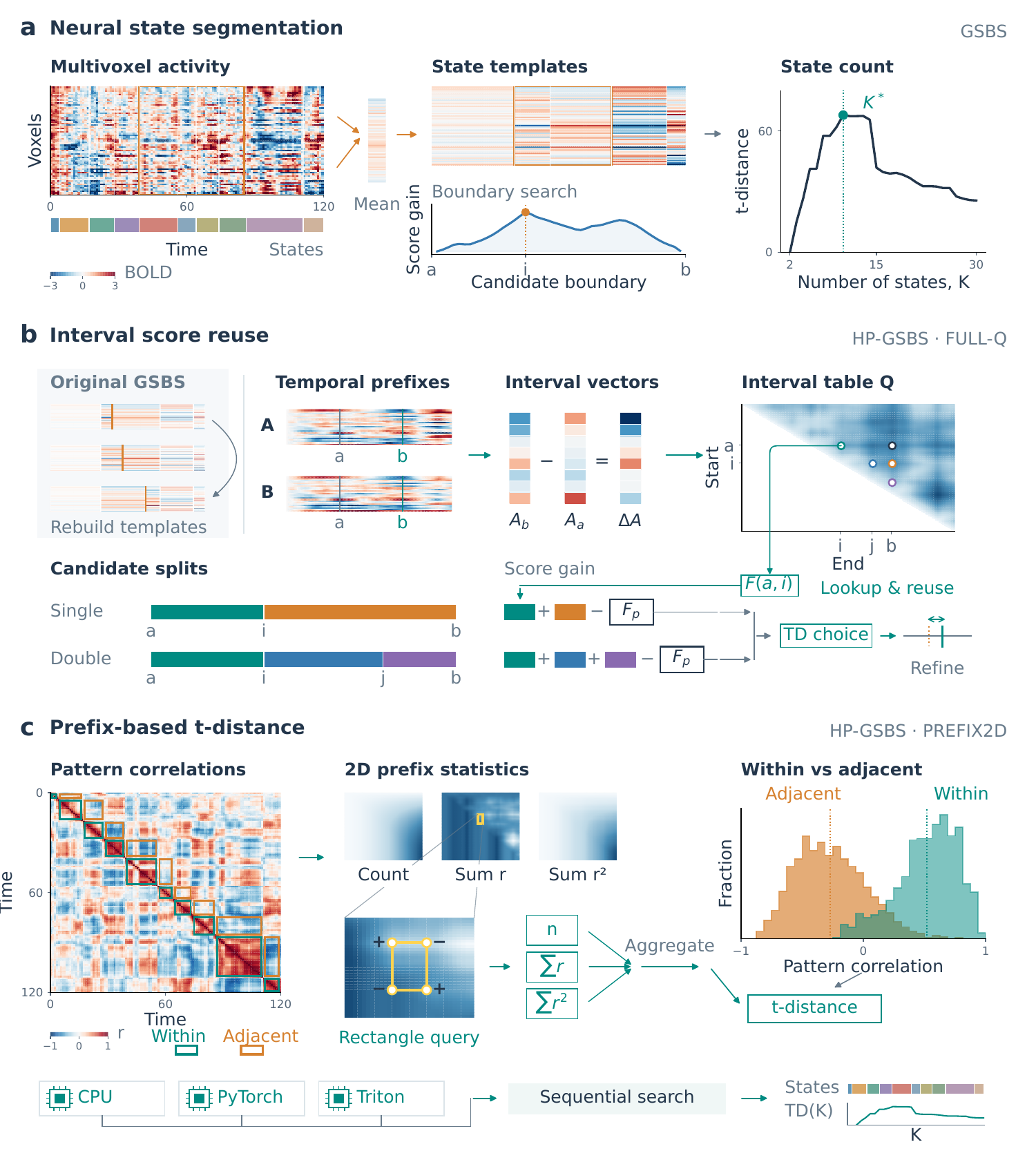}{\figOneCaption}{fig:architecture}

\subsection{Additive interval statistics}\label{sec:intervals}
The repeated calculation of correlations between observations and state templates is a major source of computational cost during candidate search. We can reorganize this calculation by exploiting the fact that Pearson correlation is a normalized inner product between spatially centered activity patterns. Let $H=I-\mathbf{1}\mathbf{1}^\top/P$ be the spatial centering operator. Applying $H$ to an observation $x_t$ removes its mean across features, giving $c_t=Hx_t$. Its sample-standardized pattern is
\[
z_t=\frac{\sqrt{P-1}\,c_t}{\|c_t\|_2}.
\]
The recorded observations remain fixed throughout the search, so their centered and standardized patterns can be computed once. To obtain the sum of either set of patterns over any interval without revisiting its individual time points, we construct two temporal prefix sums:
\begin{equation}
A_b=\sum_{t<b}c_t,\qquad B_b=\sum_{t<b}z_t.
\label{eq:prefix}
\end{equation}
With these prefixes, the sums over an interval $[a,b)$ are given by $\Delta A_{ab}=A_b-A_a$ and $\Delta B_{ab}=B_b-B_a$. The former determines the state template, while the latter summarizes the standardized observations to be compared with that template.

Centering the mean pattern gives $Hm_{ab}=\Delta A_{ab}/(b-a)$. Since Pearson correlation depends on the direction of the centered template, the factor $1/(b-a)$ cancels when the template is normalized. Substituting the normalized template into Eq.~\eqref{eq:objective} and collecting the contributions from each interval yields
\begin{equation}
\begin{split}
J(\mathcal S)&=\sum_{[a,b)\in\mathcal S}F(a,b),\\
F(a,b)&=\frac{\Delta B_{ab}^{\mathsf T}\Delta A_{ab}}
{T\sqrt{P-1}\,\|\Delta A_{ab}\|_2}.
\end{split}
\label{eq:interval}
\end{equation}
Equation~\eqref{eq:interval} expresses the contribution of an entire interval using two differences of precomputed vectors, a dot product, and a vector norm. The arithmetic required for this interval score scales with the feature count $P$, independently of the length of the interval. Moreover, the contribution $F(a,b)$ depends only on the observations between its endpoints, regardless of how the rest of the recording is segmented. The identity holds for state means with nonzero spatial variance, for which Pearson correlation is defined. State templates with zero spatial variance are handled using the reference calculation (Online Resource~1, Section~S1).

Consider a state occupying the interval $[a,b)$. Inserting a boundary at $i$ replaces this state by two shorter states and leaves all other states unchanged. The resulting gain in the matching score is therefore
\begin{equation}
\Delta_1(i)=F(a,i)+F(i,b)-F(a,b),\label{eq:single}
\end{equation}
The same principle applies when two boundaries are inserted simultaneously. For $i<j$, the contribution of the original state is replaced by those of three states:
\begin{equation}
\Delta_2(i,j)=F(a,i)+F(i,j)+F(j,b)-F(a,b).\label{eq:double}
\end{equation}
All other interval contributions can be reused. For a local search block of length $L_b$, multiplying the score by $T/L_b$ reproduces the normalization used in the reference implementation.

There are two ways to exploit Eq.~\eqref{eq:interval}. When sufficient memory is available, we calculate the score of each valid interval in advance and store it in a table $Q$, where $Q_{a,b}=F(a,b)$. We refer to this strategy as Full-Q. Building the table requires $O(T^2P)$ time and $O(T^2)$ storage, but each subsequent candidate evaluation uses only a fixed number of table accesses and scalar operations. For very long recordings, storing all interval scores may require substantial memory. An alternative, which we call raw-prefix scoring, retains the prefix vectors and computes each interval score as it is needed. This strategy uses $O(P)$ operations per interval and avoids storing the quadratic interval-score table.

\subsection{T-distance from sufficient statistics}\label{sec:td}
The matching objective identifies promising boundary locations. To evaluate a resulting segmentation and select a suitable number of states, GSBS also uses t-distance. The underlying idea is that activity patterns should be relatively similar within a state and less similar across neighboring states. We quantify this property using the spatial Pearson correlation $r_{ij}$ between the observations at time points $i$ and $j$.

Let $s(i)$ denote the state to which time point $i$ belongs, and let $d$ denote the temporal exclusion parameter $d_{\min}$. Neural recordings often exhibit temporal autocorrelation, so patterns measured close together in time may be highly similar. Increasing $d$ reduces the influence of these short-lag correlations on the comparison. We retain upper-triangular pairs satisfying $j-i\geq d$. The correlations are collected into two distributions: those for pairs within the same state, and those for pairs in consecutive states:
\begin{align}
W&=\{r_{ij}:j-i\geq d,\ s(i)=s(j)\},\\
A&=\{r_{ij}:j-i\geq d,\ s(j)=s(i)+1\}.
\end{align}
Each eligible pair is counted once, and the adjacent-state distribution contains only pairs drawn from consecutive states.

Computing t-distance requires only the mean, variance, and number of values in each of these distributions. For either distribution, these quantities can be obtained from its count $n$, sum $S$, and squared sum $R$:
\begin{equation}
\mu=S/n,\qquad v=\frac{R-S^2/n}{n-1}.
\label{eq:moments}
\end{equation}
The GSBS partition criterion is the Welch statistic \citep{welch1947},
\begin{equation}
\mathrm{TD}=\frac{\mu_W-\mu_A}{\sqrt{v_W/n_W+v_A/n_A}}.
\label{eq:td}
\end{equation}
It measures the difference between the two correlation means relative to their variability and sample counts. GSBS uses this standardized contrast as a descriptive segmentation score to compare candidate partitions and select the number of states.

We can obtain the six required summary statistics efficiently by exploiting the geometry of the time-by-time correlation matrix (Fig.~\ref{fig:architecture}c). Because each state occupies a contiguous time interval, pairs of time points within a state lie in a square block along the matrix diagonal. Pairs belonging to consecutive states lie in the rectangular blocks between these diagonal blocks. With ineligible pairs masked out, the required counts, sums, and squared sums can therefore be assembled by summing over these regions.

To retrieve such sums without visiting individual correlations each time the segmentation changes, we construct three zero-padded two-dimensional prefix tables: one for the indicator of eligible pairs, one for the masked correlations, and one for their squares. Each prefix entry $\Pi_{u,v}$ stores the sum over rows preceding $u$ and columns preceding $v$. The sum over a rectangle with row interval $[a,b)$ and column interval $[c,e)$ is then
\[
\Pi_{b,e}-\Pi_{a,e}-\Pi_{b,c}+\Pi_{a,c}.
\]
The rectangle sum requires four table accesses, regardless of how many time-point pairs the rectangle contains. We refer to this method as Prefix2D. For a segmentation with $K$ states, the within-state and adjacent-state statistics are obtained by querying $K$ diagonal blocks and $K-1$ rectangles linking neighboring states. Consequently, once the interval endpoints are available, evaluating t-distance requires $O(K)$ rectangle queries, while extracting the endpoints from the stored boundary vector requires $O(T)$ time. Full-Q and Prefix2D thus address separate stages of the search: the former supplies state-template scores, and the latter supplies summaries of correlations between time points.

Pairwise correlations are calculated from spatially standardized observations, with residual row means removed to match the covariance convention of GSBS. The standard calculation uses double-precision floating-point arithmetic (FP64). The reference small-sample convention is retained, and numerically sensitive variance estimates or close proposal comparisons are reevaluated using the original expressions. For long recordings with relatively few features, Online Resource~1 (Section~S2) derives an alternative expression for the correlation statistics using feature-vector sums and sums of their outer products within each state. This feature-moment formulation replaces dense time-by-time tables with summaries in feature space.

\subsection{Candidate search and sequential refinement}
When GSBS searches for two boundaries within a state, it must evaluate many possible pairs of boundary locations. Storing the scores of all these pairs requires an array whose size grows quadratically with the length of the searched interval. However, to select the best candidate, the algorithm only needs the current highest score and its corresponding boundary locations. We therefore visit the eligible candidates in the reference order and update the best candidate as the search proceeds. This procedure is referred to as streaming reduction. With Full-Q, each candidate is evaluated by combining interval scores retrieved from the table. With raw-prefix scoring, candidate scores are evaluated in chunks of at most $q$ candidates. Candidate enumeration, selection of the first maximum when scores are tied, and the subsequent single-versus-double comparison follow the reference procedure. Streaming reduces the storage needed for candidate scores while retaining the original candidate domain and tie-breaking rules.

Once new boundaries have been accepted, GSBS revisits existing boundary locations to improve the fit. Because moving a boundary changes only the intervals on either side of it, the interval statistics from Section~\ref{sec:intervals} can also be used during refinement. HP-GSBS removes and reinserts each boundary in sequence, evaluating candidate locations within the specified refinement radius (one time point for \texttt{finetune=1}). Boundary strength is defined as $1-\operatorname{corr}(m_k,m_{k+1})$, the Pearson correlation distance between the templates of neighboring states. With \texttt{finetune\_order=True}, boundaries are refined from the weakest to the strongest, following the GSBS ordering. Each move updates the current segmentation before the next boundary is refined, preserving the dependence of later moves on earlier ones.

A related opportunity for reuse arises when a boundary update leaves a local search block unchanged. If both the segmentation within the block and its set of eligible candidates remain the same, its previously identified best candidate remains valid. We retain the best candidate for each block in an active-block cache. Boundary insertion or refinement invalidates the cached results of affected blocks, which are searched again when needed.

The output retains the complete boundary history and the reference state-count indexing. Because the statewise algorithm can insert two boundaries in one iteration, two neighboring state-count indices may refer to the same accepted partition. A final double insertion can also produce a partition with $K_{\max}+1$ states. Preserving these indexing conventions allows direct comparison with the reference output and subsequent inspection of the complete t-distance curve.

Table~\ref{tab:costs} summarizes the computational costs of these interval-scoring, t-distance, and candidate-search strategies.

\begin{table*}[t]
\caption{Principal computational costs of the reference and HP-GSBS calculations described in Sections~2.2--2.4. $L$ is the length of the searched block, $q$ the number of candidates processed in one raw-prefix chunk, $K$ the current state count, and $d=d_{\min}$ the temporal exclusion parameter. Complete-fit runtime also depends on the number of eligible candidates and accepted updates.}\label{tab:costs}
\centering\small
\begin{tabularx}{\textwidth}{@{}>{\raggedright\arraybackslash}p{.27\textwidth}*{2}{>{\raggedright\arraybackslash}X}@{}}\toprule
Operation & Reference calculation & HP-GSBS calculation \\\midrule
One candidate score & Reconstruct and normalize state templates; $O(TP)$ & Full-Q lookup: $O(1)$; raw-prefix score: $O(P)$ \\
Interval precomputation & Repeated during candidate evaluation & $O(T^2P)$ time; $O(T^2)$ Full-Q storage \\
Double-candidate score storage & Store $O(L^2)$ candidate scores & Retain the current best candidate; bounded $q\times P$ workspace in prefix mode \\
Fixed-partition t-distance & Scan the $O(T^2)$ time-by-time structure & $O(K)$ Prefix2D queries after identifying endpoints \\
Low-memory t-distance & Dense time-by-time correlation matrix & Feature moments: $O(TP^2+KP^2+TdP)$ arithmetic \\
Complete boundary history & $O(K_{\max}T)$ output & Retain the same $O(K_{\max}T)$ output \\
\botrule\end{tabularx}
\end{table*}

\subsection{CPU and GPU implementations}
The reformulations above separate the computation into two stages. First, reusable interval scores and correlation statistics are constructed from the recorded data. These calculations consist largely of regular numerical operations that can be executed in parallel. Second, the statistics are used to select and refine boundaries, a process in which each decision can depend on previous updates to the segmentation. We implement the first stage using three numerical backends and share the sequential search procedure across them.

The CPU backend combines NumPy and SciPy array operations with Numba-compiled search kernels \citep{numpy2020,scipy2020,numba2015}. Its Full-Q implementation computes interval scores in bounded tiles and combines score evaluation with running candidate reductions. The PyTorch CUDA backend uses batched matrix products and prefix operations \citep{pytorch2019}. The Triton backend computes interval scores in blocks on the GPU, combining prefix differences, dot products, and norms within a single kernel \citep{triton2019}. Both GPU backends transfer the reusable statistics to the common CPU search controller, which performs candidate selection, t-distance comparisons, and sequential refinement in the same order as the CPU backend.

Performance therefore depends on how much time a particular analysis spends constructing the statistics, transferring data, and performing sequential search. The standard configuration combines FP64 Full-Q, Prefix2D, and streaming candidate search. The optional CPU low-memory mode combines raw-prefix scoring with the feature-moment formulation of t-distance. Experimental options include single-precision arithmetic (FP32) and on-GPU candidate reduction, which identifies the highest-scoring candidate directly on the device for selected candidate sets (Online Resource~1, Section~S3).

\subsection{Software interface and execution modes}
HP-GSBS accepts prepared time-by-feature arrays and separates the scientific configuration of a segmentation from the way the computation is executed. The scientific parameters include the maximum state count, temporal exclusion, statewise detection, block size, and boundary refinement. Execution options specify the backend, resource limits, and whether scheduling prioritizes the latency of an individual fit or the throughput of a recording collection. For example, a fit on the CPU can be requested using \texttt{fit(x, GSBSConfig(kmax=60, dmin=1), backend="cpu-fast")}. The result contains the selected state count, boundary arrays at every evaluated state-count index, the t-distance curve, state patterns, and boundary strengths. These outputs can be saved in a NumPy archive (NPZ), with version information for later comparison and visualization.

Studies involving multiple participants, brain regions, or sessions often require fitting many independent recordings. HP-GSBS supports this workflow through independent CPU workers or compatible GPU batches. A command-line interface reads a list of analysis tasks and records their inputs and configurations. Each completed segmentation is saved separately. If a batch analysis is interrupted, completed results can be reused when the input, configuration, and software version match the requested analysis. This allows researchers to resume large studies and examine the segmentation of each recording independently. Installation instructions, result schemas, and executable workflow examples are provided in the software repository and Online Resource~1 (Section~S3).

\subsection{Evaluation design and datasets}\label{sec:evaluation}
Our evaluation addresses two main questions: whether HP-GSBS reproduces the results of the reference implementation while reducing its computational cost, and how the resulting implementation performs on realistic neural recordings and collections of recordings. We examine numerical equivalence, paired complete-fit runtime, agreement on real EEG data, memory and throughput scaling, and a cohort-level fMRI application (Table~\ref{tab:datasets}). HP-GSBS 0.2.0 is compared with the unmodified \texttt{statesegmentation} 0.0.6 implementation. Boundary histories and selected state counts must agree exactly. For t-distance, state patterns, and boundary strengths, agreement is evaluated using absolute and relative tolerances of $10^{-8}$, together with the numerical criteria detailed in Online Resource~1 (Section~S4).

\begin{table*}[t]
\caption{Datasets and the questions addressed by each evaluation. $T\times P$ denotes the number of time points by the number of features. Section~\ref{sec:evaluation} and Online Resource~1 specify the numerical agreement criteria, scientific configurations, and timing protocols.}\label{tab:datasets}
\centering\small
\begin{tabularx}{\textwidth}{@{}>{\raggedright\arraybackslash}p{.20\textwidth}>{\raggedright\arraybackslash}p{.23\textwidth}>{\raggedright\arraybackslash}p{.19\textwidth}>{\raggedright\arraybackslash}X@{}}\toprule
Evaluation & Input & Configuration or scope & Principal evidence \\\midrule
Controlled comparison & Seven configurations; two seeds; $T=256$--768, $P=8$--64 & $K_{\max}=10$--40; 14 paired cases & Paired complete-fit speedups and numerical agreement across 168 optimized outputs \\
Illustrative signal & $80\times16$; four generating states & Seed 19; noise SD 0.05 & Recovery of known boundaries and agreement of search decisions \\
BBBD EEG & Four matrices; $T=1336$--1857; $P=32$ or 64 & 5.12 Hz; $K_{\max}=121$--172 & Numerical agreement for complete human EEG recordings \\
Resource measurements & Successive EEG search optimizations; long inputs; recording collections & Fixed configurations per series & Contributions of individual optimizations, peak memory, and throughput \\
Sherlock fMRI & 16 participants; each $946\times4230$ & TR 1.5 s; $K_{\max}=60$; 17 CPU fits & Group and individual segmentations and intersubject boundary alignment \\
\botrule\end{tabularx}
\end{table*}

\paragraph{Controlled benchmark.}
The cost of a GSBS analysis depends on the length of the recording, the number of features, and the number of states explored during the search. To examine the effect of each factor, we vary it while holding the other two fixed. The seven unique configurations vary $T\in\{256,512,768\}$, $P\in\{8,32,64\}$, or $K_{\max}\in\{10,20,40\}$ around the shared central configuration $(512,32,20)$. For each configuration, we generate inputs using seeds 20261017 and 20261018. Each input comprises eight contiguous states with Gaussian spatial means and additive Gaussian noise of standard deviation 0.8. Feature-count comparisons use nested subsets of features, and comparisons across maximum state counts reuse the same inputs. Thus, the 14 configuration--seed combinations contain ten distinct matrices. Each combination has one timed official fit and, for each optimized backend, one warm-up fit and three timed fits. We compare the segmentation results from all 168 optimized runs, including the warm-up fits, with the corresponding reference results.

To make runtime comparisons directly interpretable, we time complete fits on the same hardware using identical inputs and scientific configurations. Measurements use an AMD Ryzen 7 7840H CPU and an NVIDIA RTX 4050 Laptop GPU (6 GB), Ubuntu 22.04 under WSL2, Python 3.10.12, NumPy 2.2.6, SciPy 1.15.3, Numba 0.68.0, PyTorch 2.11.0 with CUDA 12.8, and Triton 3.6.0. Numerical-library thread counts are set to one and TF32 is disabled. Backend order is rotated across cases, and GPU calls are synchronized immediately before and after timing. Complete-fit times include construction, computation, and required transfers; validation and file operations occur outside the timed interval. For each input, we divide the official runtime by the median runtime of the three timed fits for a given optimized backend. Configuration summaries report the median of the two resulting same-input ratios and their observed range. We also examine the runtime contribution of successive optimizations, peak memory use, and the rate at which collections of recordings can be processed (Online Resource~1, Sections~S4--S6).

\paragraph{EEG.}
The controlled benchmarks test agreement under systematically varied conditions. To examine whether agreement is maintained in realistic neural recordings and in searches involving more than one hundred states, we also segment four preprocessed EEG recordings from BBBD \citep{bbbd2026}. The recordings contain 1,336--1,857 time points across either 32 frontal channels or all 64 channels. Preprocessing included cropping around stimulus events, masking artifacts, averaging consecutive nonoverlapping groups of 25 samples to reduce the sampling rate from 128 to 5.12 Hz, and standardizing each channel within a recording using the sample standard deviation of valid time bins. The stored float32 inputs are converted to FP64 for fitting. All recordings use $d_{\min}=5$, statewise detection, a one-sample refinement radius, strength-based refinement ordering, and a block size of 50, with recording-specific $K_{\max}$ values (Table~\ref{tab:eeg}). For Recording 1, the CPU, PyTorch CUDA, and Triton results are compared directly with the official \texttt{statesegmentation} output; the corresponding official fit took approximately 1.7 hours (Online Resource~1, Table~S8). For Recordings 2--4, we reuse saved CPU results previously verified against the official implementation as the reference for the GPU comparisons. The reproducibility archive provides the recording identifiers, preprocessing details, and corresponding reference results (Online Resource~1, Section~S4).

\paragraph{Naturalistic fMRI.}
A further aim of HP-GSBS is to make it practical to segment high-dimensional recordings from multiple individuals and compare the resulting state boundaries. We illustrate this application using the public Sherlock movie-viewing dataset, specifically the angular-gyrus array distributed by \citet{baldassano2020sherlock} and derived from \citet{chen2017shared}. The array contains 16 participants, 946 time points at a repetition time (TR) of 1.5 s, and 4,230 voxels, covering 23.65 min of the movie. We retain the supplied preprocessing and convert each matrix to FP64. We first segment the arithmetic mean response across participants and then segment each participant's response independently, allowing us to inspect both the group-level segmentation and the individual estimates. All 17 fits use the CPU Full-Q/Prefix2D path, $K_{\max}=60$, $d_{\min}=1$, statewise detection, a one-sample refinement radius, strength-based ordering, and a block size of 50. Each complete fit is timed once on the same CPU platform with single-thread numerical libraries.

Because all participants viewed the same movie, we ask whether their estimated neural state transitions tend to occur at similar times. The number of detected boundaries can differ between individuals, so we quantify alignment using a pairwise $F_1$ score based on one-to-one boundary matching. Boundaries are considered eligible to match when they occur within $\pm3$ TR ($\pm4.5$ s), and each boundary can be matched at most once. If participants $i$ and $j$ have $n_i$ and $n_j$ boundaries and $M_{ij}$ matched pairs, their score is $F_{1,ij}=2M_{ij}/(n_i+n_j)$. The cohort statistic is the mean of these scores across all 120 participant pairs.

Some boundaries are expected to coincide by chance, especially when the number of boundaries is large. To estimate this chance alignment while retaining each individual's boundary count and circular interboundary spacing, we independently shift each participant's boundary sequence by a uniformly sampled nonzero offset on a circular time axis. Each of 1,000 permutations produces a new cohort-mean $F_1$. After shifting, matching is based on absolute temporal distance along the linear recording timeline. The one-sided Monte Carlo probability is $(1+\#\{F_1^{\mathrm{null}}\geq F_1^{\mathrm{observed}}\})/1001$, using seed 20261020. Online Resource~1, Section~S8 provides the complete outputs and implementation details.

We generated figures from the numerical arrays using Matplotlib \citep{matplotlib2007}. ChatGPT and Codex (OpenAI) assisted with software engineering and plotting scripts; the full disclosure is provided in the Declaration of generative AI use.

\section{Results}\label{sec:results}
\subsection{Numerical equivalence across synthetic and EEG data}
We first examined whether the two mathematical reformulations and the resulting search optimizations preserve the segmentations produced by the reference implementation. Across all seven controlled benchmark configurations, both random seeds, and all three backends, the 168 optimized runs reproduced the reference boundary histories and selected state counts exactly. Agreement was maintained at every evaluated stage of the search, and the resulting state patterns and boundary strengths were also identical. The maximum absolute difference in t-distance was $1.88\times10^{-11}$, well below the absolute tolerance of $10^{-8}$. Thus, the optimized calculations preserved the greedy search trajectory throughout the controlled benchmarks.

To illustrate this agreement for a recording with known state boundaries, we generated a signal containing four states and 80 observations across 16 features (Fig.~\ref{fig:consistency}). The generating boundaries occur at samples 20, 40, and 60, and Gaussian noise has a standard deviation of 0.05. All four implementations identify these three boundaries and select four states. Their complete t-distance curves overlap at the scale of the plot, with maximum absolute differences from the reference of $1.51\times10^{-11}$ for CPU and $6.93\times10^{-12}$ for both GPU backends. Step-by-step traces of boundary search and refinement also agree with the reference at every recorded decision. The additional simulations in Table~S12 of Online Resource~1 show that higher noise reduces recovery of the generating boundaries, while all implementations agree on the resulting segmentations.

\mainfigure{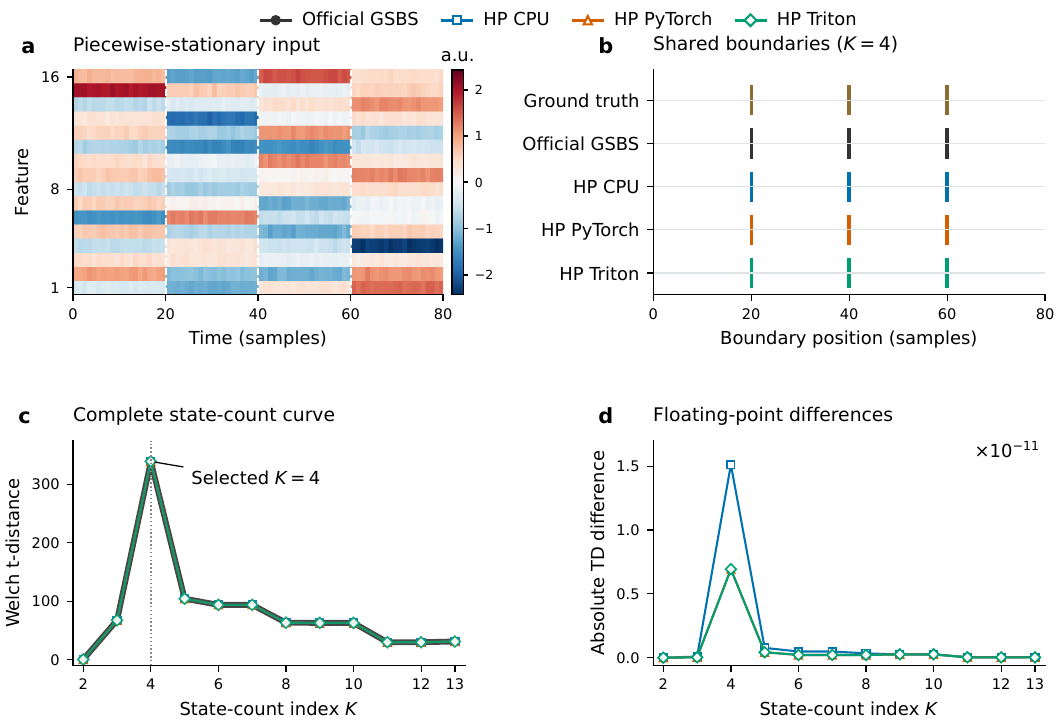}{\figTwoCaption}{fig:consistency}

We next tested whether this agreement extends to human EEG recordings with realistic, nonstationary noise and searches involving many more states. Across the four BBBD recordings, the selected state counts range from 120 to 173, corresponding to mean state durations of approximately 2.0--3.0 s at 5.12 Hz (Table~\ref{tab:eeg}). CPU, PyTorch CUDA, and Triton reproduce the corresponding reference boundary histories exactly, and the state patterns and boundary strengths agree within the prescribed numerical tolerances. The maximum absolute t-distance differences range from $1.85\times10^{-13}$ to $3.13\times10^{-13}$.

Figure~\ref{fig:eegapplication} shows the segmentation of the complete 64-channel EEG recording for Recording 1. The official implementation and all three HP-GSBS backends select 126 states with identical boundary locations and a maximum absolute t-distance difference of $2.43\times10^{-13}$. In the enlarged 20 s window (Fig.~\ref{fig:eegapplication}b--c), the estimated boundaries separate periods with different multichannel activity patterns. The corresponding time-by-time correlation matrix contains groups of consecutive time points with relatively similar spatial patterns, consistent with the identified state intervals. Across the full recording, the t-distance curves from all implementations overlap at the scale of the plot and reach their highest value at $K=126$ (Fig.~\ref{fig:eegapplication}d).

\mainfigure{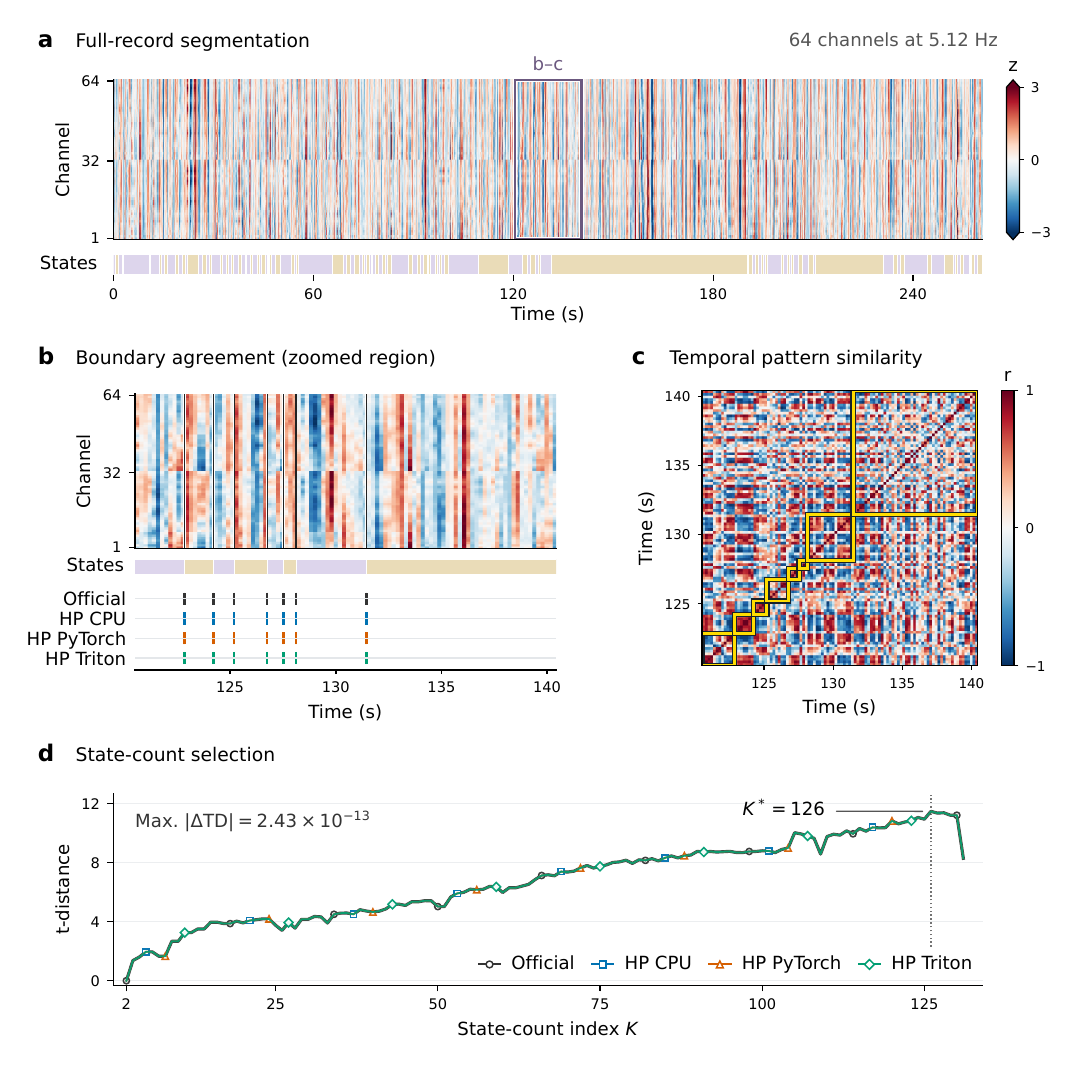}{\figThreeCaption}{fig:eegapplication}
\begin{table*}[t]
\caption{Numerical agreement across HP-GSBS CPU, PyTorch CUDA, and Triton backends for four BBBD EEG recordings, using FP64. The complete boundary histories and selected state counts agree for all recordings; the table reports the largest absolute t-distance difference across the compared outputs. All recordings use $d_{\min}=5$, statewise detection, \texttt{finetune=1}, strength-based refinement ordering, and a block size of 50. Recording 1 is compared directly with the official implementation; Recordings 2--4 use CPU reference outputs previously validated against the official implementation. The reference double-insertion convention permits the $K_{\max}+1$ index.}\label{tab:eeg}
\centering\small
\begin{tabular}{@{}crrrrc@{}}\toprule
Recording & $T\times P$ & $K_{\max}$ & Selected $K$ & Maximum $|\Delta\mathrm{TD}|$ & Identical boundary histories \\\midrule
1 & $1336\times64$ & 131 & 126 & $2.43\times10^{-13}$ & Yes \\
2 & $1857\times32$ & 121 & 120 & $3.13\times10^{-13}$ & Yes \\
3 & $1754\times64$ & 172 & 173 & $2.77\times10^{-13}$ & Yes \\
4 & $1374\times64$ & 135 & 131 & $1.85\times10^{-13}$ & Yes \\
\botrule\end{tabular}
\end{table*}

\subsection{Speedups, computational reuse, and resource scaling}
Having established numerical agreement with the reference implementation, we next examined how much runtime is saved by HP-GSBS. Across the controlled benchmark configurations and backends, median speedups relative to the official implementation on the same inputs range from 154- to 1,730-fold (Fig.~\ref{fig:performance}). For the central configuration, the official implementation takes 64.7 s, compared with 0.142 s for HP-GSBS on CPU, 0.081 s for PyTorch CUDA, and 0.099 s for Triton. The median paired speedup ratios are 454, 795, and 675, respectively.

The computational advantage becomes particularly pronounced as recording length increases (Fig.~\ref{fig:performance}a,d). This pattern is consistent with the growing cost of repeatedly constructing state templates in the reference implementation. At $T=768$, HP-GSBS on CPU reaches a 770-fold speedup and PyTorch CUDA reaches a 1,730-fold speedup. Across the feature-count and search-depth comparisons at $T=512$, all optimized backends complete fits in 0.065--0.165 s (Fig.~\ref{fig:performance}b--c). The large gains on the same CPU demonstrate the benefit of reorganizing and reusing computations. At $T=768$, PyTorch CUDA provides a further roughly twofold reduction in runtime relative to the optimized CPU, consistent with the suitability of dense interval precomputation for GPU execution. Complete benchmark tables and additional larger workloads, including inputs with up to 2,100 time points and a 1,336-sample EEG recording, are reported in Online Resource~1 (Sections~S5--S6).

\mainfigure[!t]{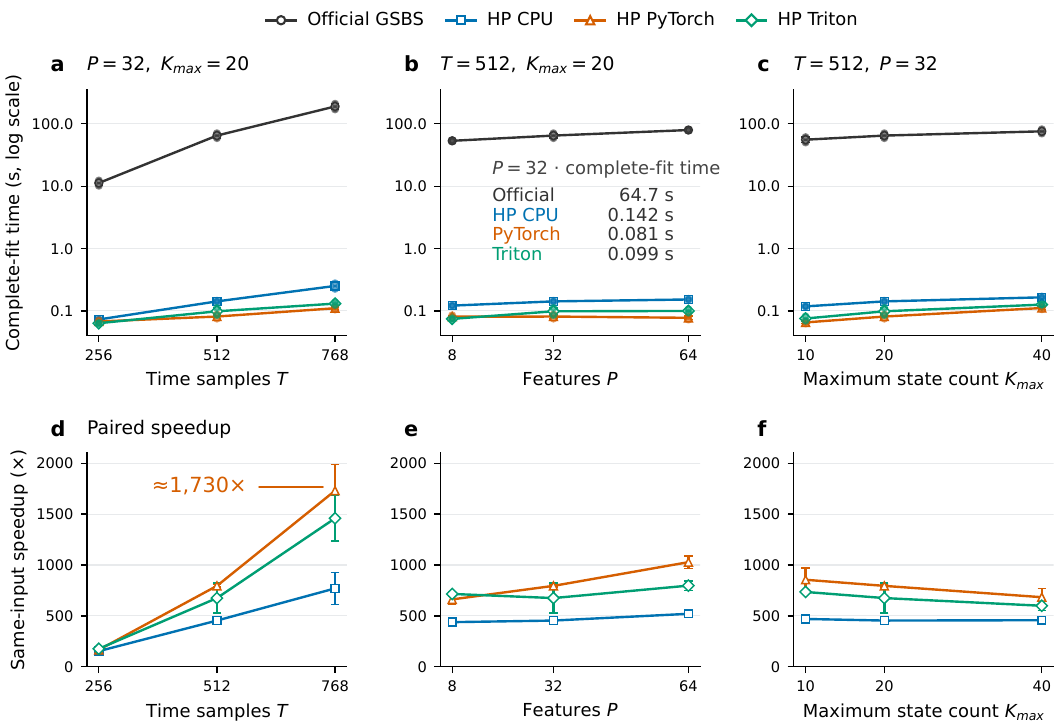}{\figFourCaption}{fig:performance}

Once Full-Q has made interval scores inexpensive to retrieve, additional costs arise from storing candidate scores, refining boundaries, and repeatedly searching unchanged blocks. To examine the contribution of each subsequent optimization, we conducted a sequential CPU ablation using the $1{,}336\times64$ EEG recording with $K_{\max}=131$ (Fig.~\ref{fig:practical}a). The starting implementation already precomputes the interval scores. Replacing the stored candidate-score arrays with streaming reduction decreases the median complete-fit time from 3.364 to 1.803 s. Evaluating refinement candidates within their specified local windows further decreases it to 1.601 s, and caching the best candidates in unchanged search blocks brings it to 1.429 s. All four configurations produce identical segmentations. These results show that each optimization reduces a distinct source of search-stage cost within the precomputed interval-score framework.

The preceding benchmarks use precomputed tables whose storage requirements grow quadratically with the number of time points. For long recordings, this motivates computing interval scores from their temporal prefixes as needed and summarizing correlations in feature space. We evaluated this low-memory formulation on a single-boundary search, whose candidate count grows linearly with interval length. This configuration separates the effect of memory-efficient scoring from the quadratic enumeration of double-boundary candidates. All fits in this series used a block size of zero, omitted boundary refinement, and explored up to $K_{\max}=16$ states (Fig.~\ref{fig:practical}b). On a $5{,}000\times8$ input, the low-memory formulation reduced peak resident set size (RSS) from 1,598.0 to 190.6 MiB and fit time from 9.754 to 0.110 s relative to the Full-Q configuration with stored candidate scores. With both recording length and feature count increased to $50{,}000\times64$, the fit took 1.444 s and reached a peak RSS of 469.5 MiB. A single dense double-precision correlation matrix for this recording would occupy 20 GB. We therefore evaluated the t-distance of the selected and final partitions using an independent blockwise calculation of pairwise correlations; the maximum absolute difference was $1.15\times10^{-14}$.

In studies involving many independent recordings, the number of fits that can be completed within a given time is also important. We examined throughput when distributing recordings among CPU workers or batching compatible inputs on the GPU (Fig.~\ref{fig:practical}c,d). For short synthetic inputs, CPU throughput increases from 274 fits/minute with one worker to 1,767 fits/minute with eight workers. PyTorch throughput increases from 647 fits/minute with a batch size of one to 1,418 fits/minute with a batch size of eight. For repeated EEG tasks requiring many more state updates, eight CPU workers complete 155 fits/minute, compared with approximately 40--45 fits/minute for the GPU backends. This difference is consistent with the structure of the computation: batching is useful when constructing the reusable statistics accounts for a substantial part of the fit, whereas searches over many state counts spend more time in the sequential controller. Distributing the latter analyses across independent CPU workers can therefore yield higher overall throughput.

\mainfigure[!t]{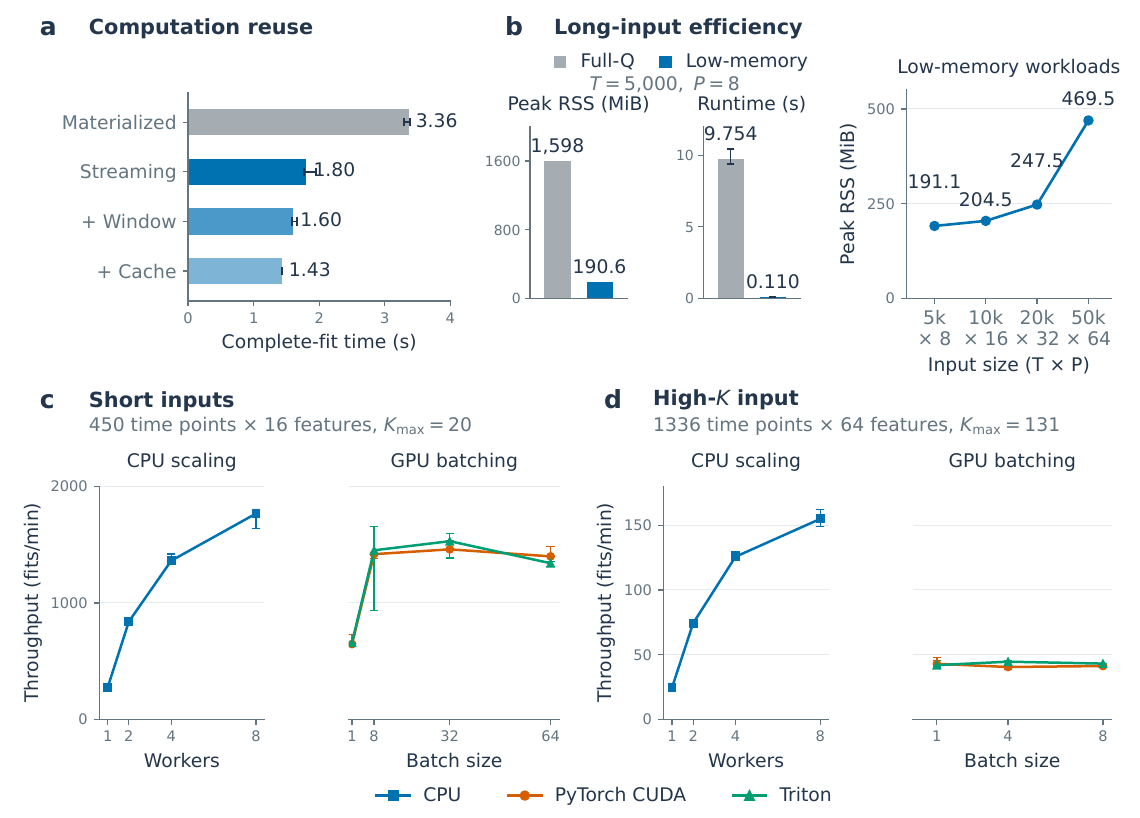}{\figFiveCaption}{fig:practical}

\subsection{Naturalistic fMRI segmentation and intersubject boundary alignment}
We next used HP-GSBS to examine the temporal alignment of neural state boundaries across individuals viewing the same naturalistic stimulus. Applying the CPU implementation to the Sherlock angular-gyrus dataset (4,230 voxels per participant, 946 TRs, 23.65 min) takes 13.56 s for the group-mean response and 11.88--13.15 s for each of the 16 individual responses (median 12.38 s; Fig.~\ref{fig:fmri}). All 17 fits were completed in 213 s on the CPU, illustrating the feasibility of voxel-level cohort segmentation on laptop-class hardware.

The t-distance curves and their selected state counts vary across individuals (Fig.~\ref{fig:fmri}b--c). Within the evaluated search range, the selected state counts span 19 to 61 (median 56.5). Nine participants have their highest t-distance below 60, while the group response and seven participants select 60 or 61 states. The configured maximum is $K_{\max}=60$, with the original double-insertion convention allowing index 61. Thus, the fitted segmentations differ in temporal granularity, and several state-count selections lie at the upper end of the evaluated range. The latter observation motivates extending the range when studying the temporal scales of individual responses.

Despite the variability in the selected state counts, estimated transitions tend to occur at similar moments across viewers (Fig.~\ref{fig:fmri}d). With a matching tolerance of $\pm3$ TR ($\pm4.5$ s), the observed mean pairwise $F_1$ was 0.414, exceeding the values from all 1,000 circular-shift permutations (null mean 0.325, central 95\% interval [0.315, 0.335], difference 0.089, $p=0.001$). The individual GSBS segmentations therefore reveal a shared temporal component in the estimated angular-gyrus state boundaries, alongside variability in the individual segmentations.

\mainfigure[!t]{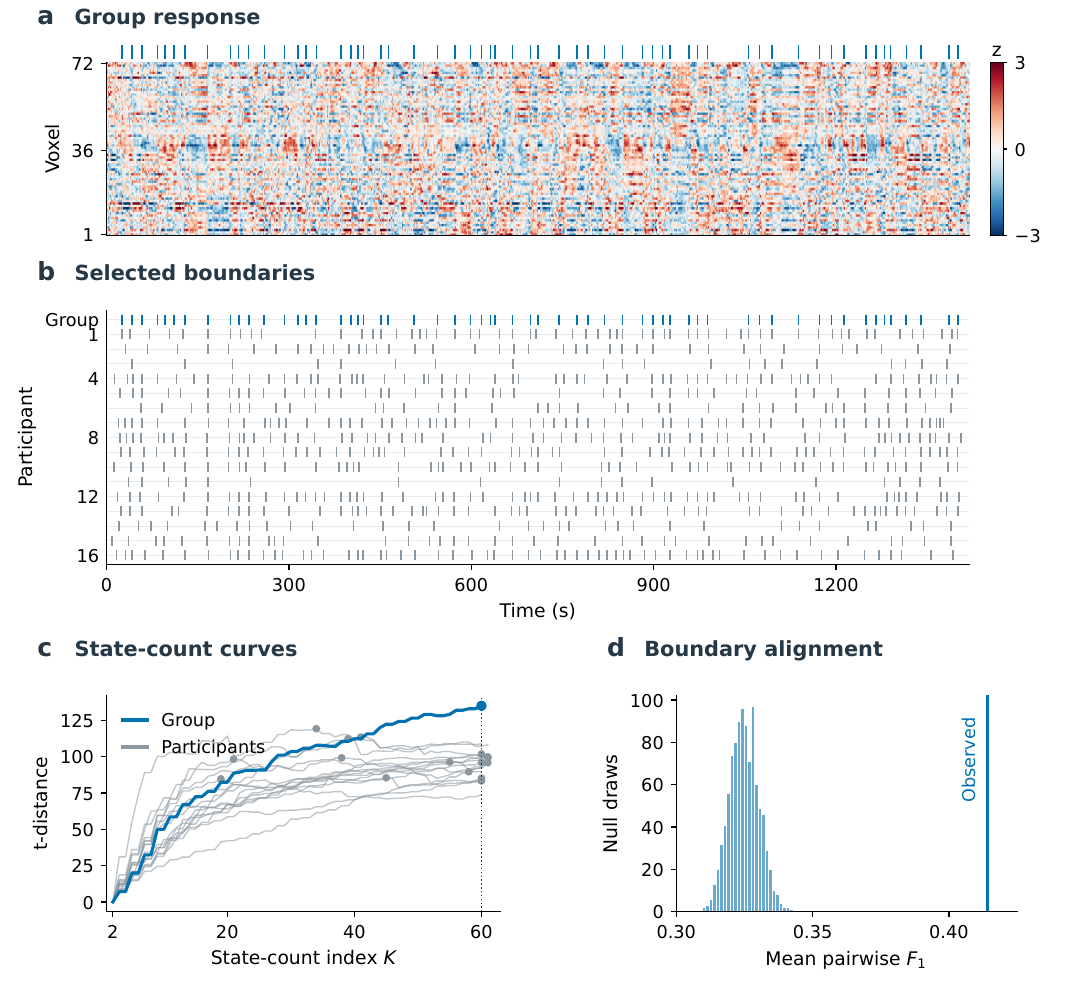}{\figSixCaption}{fig:fmri}

\section{Discussion}\label{sec:discussion}
\subsection{Computational contributions}
The central computational observation underlying HP-GSBS is the locality of a boundary update: changing a boundary alters only a small part of the segmentation. The reference implementation repeatedly evaluates calculations over much larger portions of the recording, creating opportunities to reuse previously computed quantities. By expressing the matching objective as a sum of interval-specific contributions, we make this locality explicit. Temporal prefix sums allow each interval contribution to be computed efficiently, and Full-Q allows the same contribution to be retrieved whenever the interval reappears in a candidate segmentation. The correlation statistics used for t-distance exploit a related property: the pairwise correlations are fixed by the data, while changes in the boundaries determine which of these correlations enter the two distributions. Once the prefix tables have been constructed, a segmentation can be evaluated by combining summaries of the corresponding regions of the correlation matrix. The two reformulations therefore allow a large fraction of the repeated work in GSBS to be replaced by the reuse of precomputed quantities.

The ablation results illustrate how additional search optimizations complement the interval reformulation. Streaming candidate evaluation reduces the cost of storing candidate scores; local refinement concentrates calculations within the specified search window; and caching avoids searching a block again when its local problem has remained unchanged. Numerical agreement is particularly important for a greedy algorithm because a change in a boundary decision can affect later candidates and refinement moves. Across the controlled benchmarks and EEG recordings, the optimized implementations reproduce the corresponding complete reference boundary histories, with t-distance differences below $2\times10^{-11}$, consistent with rounding and differences in the order of floating-point operations. Checking every evaluated partition, the selected state count, and the complete t-distance curve makes it possible to assess these numerical differences in relation to the actual segmentation decisions.

\subsection{Implications for scalable neuroscience analysis}
Investigating the temporal organization of neural activity across the cortex may require fitting state segmentations separately in many brain regions. Studying individual differences, comparing stimuli, or testing the sensitivity of results to analysis parameters can multiply the number of required fits. The computational improvements demonstrated here reduce minute-scale fits to subsecond calculations in the controlled benchmarks on laptop-class hardware. Independent recording-level scheduling extends this benefit to collections, and the saved boundary histories allow researchers to inspect the estimated segmentations across state counts without rerunning the same fit. Saving the analysis settings together with each segmentation also supports reanalysis and comparisons across studies.

The two empirical datasets illustrate different computational demands. The BBBD EEG recordings require searches over more than one hundred states, whereas the Sherlock fMRI application involves thousands of spatial features and multiple individual recordings. The shared analysis interface supports both types of data. The boundary alignment observed in the Sherlock dataset is consistent with previous work on shared neural event structure during narrative processing \citep{baldassano2017,chen2017shared,geerligs2022}. At the same time, estimating the temporal granularity of individual responses warrants attention to the search range: seven participants and the group response select state-count indices at or near the configured upper limit of 60. Expanding the state-count range and examining the sensitivity of boundary alignment to state count and temporal matching tolerance would help characterize the observed temporal structure more fully. With individual fits taking about 12 s in this application, HP-GSBS makes such analyses, including comparisons across cortical regions, more accessible.

\subsection{Implementation and resource trade-offs}
The most suitable computational backend depends on which stage of an analysis accounts for most of its runtime. The standard FP64 Full-Q/Prefix2D configuration used in the evaluated statewise analyses requires quadratic table storage, but allows the scores and correlation summaries needed throughout the search to be retrieved efficiently. When precomputation makes up a substantial fraction of the fit, GPU execution and batching compatible inputs can improve performance. When sequential boundary search is the main cost, or when many recordings can be processed independently, CPU workers may offer higher throughput. The results for short synthetic inputs and high-state-count EEG tasks illustrate these two scenarios. The trade-off on other systems will also depend on available memory, data-transfer costs, and hardware performance in FP64 arithmetic.

For long recordings with modest feature counts, the optional low-memory mode offers a complementary time--space trade-off. It computes interval scores from the prefixes as they are needed and summarizes pairwise correlations through feature-space statistics. The long-input benchmarks use single-boundary search without refinement, for which the number of candidate locations grows linearly with interval length. When statewise detection is enabled, the number of possible double-boundary candidates within a long interval still grows quadratically, even if the score of each candidate can be computed efficiently. In addition, retaining the complete boundary history requires $O(K_{\max}T)$ storage in all execution modes. These costs are relevant when choosing the search configuration and computational resources for a particular analysis.

\subsection{Future development}
Several aspects of the computation provide opportunities for further development. Since a boundary update changes the correlation regions associated with only part of a segmentation, maintaining and updating the corresponding statistics incrementally could further reduce the number of t-distance queries. For large candidate domains, performing selected reductions directly on the GPU could also reduce search-stage or data-transfer costs. Another direction is to examine how reduced numerical precision affects both computational throughput and the stability of boundary decisions when candidate scores are nearly equal. The released comparison scripts allow these extensions to be evaluated in terms of both computational speed and the stability of boundary choices throughout the search.

At the application level, analyses across more brain regions and a broader range of state counts would help determine how faster segmentation can advance studies of the temporal organization of neural activity. Parallel fitting across participants is especially useful when analyses must be repeated for resampling or parameter sensitivity. Recording explicit analysis configurations together with complete state-count curves will facilitate comparisons across larger cohorts and regions. More generally, the approach illustrated here may be useful for other iterative methods in neuroscience: when model updates operate on the same observations and alter only part of the model, identifying the intermediate quantities that remain unchanged can reveal opportunities for computational reuse. Expressing these quantities in a mathematically interpretable form provides a basis for computational optimization while retaining the underlying statistical procedure.

\section{Conclusions}
Much of the computational cost of Greedy State Boundary Search arises from calculations repeated as successive candidate partitions are evaluated. HP-GSBS exploits this redundancy by expressing template matching as a sum of reusable interval scores and computing Welch t-distance from precomputed correlation statistics. These mathematical reformulations, combined with efficient candidate search, local refinement, and caching, substantially reduce the cost of neural-state segmentation. Across controlled benchmarks, the CPU, PyTorch CUDA, and Triton FP64 implementations achieve 154- to 1,730-fold speedups while reproducing the complete reference boundary histories and selected state counts. Numerical agreement is also maintained across backends for the high-state-count EEG recordings. Applied to naturalistic fMRI data from 16 participants, HP-GSBS completes voxel-level group and individual segmentations in minutes on a single CPU and reveals significant temporal alignment of estimated state boundaries across viewers. By making repeated segmentation practical on standard laboratory hardware, HP-GSBS facilitates investigations of neural temporal organization across individuals, brain regions, recording modalities, and timescales.

\section*{Information Sharing Statement}
The software used in this study, HP-GSBS 0.2.0, is available under the MIT license at \url{https://github.com/JustinZHAO-05/HP-GSBS/tree/v0.2.0} and through its versioned Zenodo record \citep{hpgsbs2026}. The numerical inputs, outputs, and scripts needed to reproduce this article are provided in the expanded reproducibility archive, version 1.1.0 (\url{https://doi.org/10.5281/zenodo.23573202}; \citealp{hpgsbsrepro1102026}), corresponding to tag \texttt{v1.1.0} at \url{https://github.com/JustinZHAO-05/HP-GSBS-SoftwareX-Reproducibility}. The archive includes all four EEG analysis inputs, time maps, scientific configurations, and saved FP64 outputs; the complete Sherlock application, including the 17 segmentations and boundary-alignment statistics; and the data and scripts for all six main figures. The earlier version 1.0.1 archive preserves the original benchmark and figure materials \citep{hpgsbsrepro2026}; the all-versions DOI is \url{https://doi.org/10.5281/zenodo.23457341}.

Online Resource~1 (\path{ESM_1.pdf}) provides the mathematical derivations, complete benchmark and resource tables, EEG provenance and file mapping, and the Sherlock analysis protocol with reproduction commands. The original BBBD dataset is described by \citet{bbbd2026}, and the Sherlock angular-gyrus input is available from \citet{baldassano2020sherlock}. The derived inputs are accompanied by source attribution and preprocessing records. Reproduction code is MIT licensed; distributed data, derived figures, and documentation retain CC BY 4.0 attribution.

\section*{Statements and Declarations}
\subsection*{Funding}
This research received no specific grant from any funding agency in the public, commercial, or not-for-profit sectors.

\subsection*{Competing interests}
The authors declare that they have no competing interests.

\subsection*{Author contributions}
\textbf{Yanzhe Zhao:} Conceptualization, Methodology, Software, Formal analysis, Investigation, Validation, Data curation, Visualization, Writing---original draft, Writing---review and editing, Project administration.

\textbf{Yongzhang Ji:} Methodology, Formal analysis, Writing---review and editing.

\subsection*{Ethics approval and consent}
The study comprises secondary computational analyses of existing publicly shared human EEG and fMRI data. Participant recruitment, informed consent, and ethical oversight are described in the original dataset publications \citep{bbbd2026,chen2017shared}. The present analyses use the shared numerical recordings and their supplied identifiers.

\subsection*{Declaration of generative AI use}
During the preparation of this work, the authors used ChatGPT and Codex (OpenAI) to assist with software engineering, analysis and plotting scripts, manuscript drafting, and language editing. The authors conceived and directed the analyses, verified the numerical results, and take responsibility for the software and the manuscript.

\subsection*{Data and code availability}
Software, analysis data, numerical outputs, and reproduction materials are identified in the Information Sharing Statement.

\setlength{\bibsep}{.45em}
\bibliography{references}

\end{document}


\begin{center}
{\Large\bfseries Online Resource 1}\par\vspace{5pt}
{\normalsize Neuroinformatics}\par\vspace{5pt}
{\large HP-GSBS: Accelerating greedy state boundary search\newline with reusable interval statistics}\par\vspace{5pt}
Yanzhe Zhao$^{1,*}$, Yongzhang Ji$^2$\par
$^1$School of Future Technology, Tianjin University, Tianjin, China\par
$^2$Faculty of Psychology, Southwest University, Chongqing, China\par
$^*$Correspondence: \href{mailto:justinzhaoyz@outlook.com}{justinzhaoyz@outlook.com}
\end{center}

\section{Interval decomposition and candidate reuse}\label{sec:supp-1}
The purpose of the interval decomposition is to evaluate the GSBS matching objective without repeatedly constructing state templates and correlating them with the individual observations \citep{geerligs2021,geerligs2022}. The starting point is the normalized-inner-product representation of Pearson correlation. Let $H=I-\mathbf{1}\mathbf{1}^\top/P$ be the spatial centering operator, $c_t=Hx_t$ the centered observation, and $z_t=\sqrt{P-1}\,c_t/\|c_t\|_2$ its sample-standardized pattern. For a state interval $S=[a,b)$ of length $L=b-a$, its mean pattern is $m_S=L^{-1}\sum_{t\in S}x_t$. Denote the corresponding sample-standardized template by $z(m_S)$. Since the Pearson correlation between $x_t$ and $m_S$ is $z_t^\top z(m_S)/(P-1)$, the interval contributes
\begin{equation}
F(a,b)=\frac{1}{T(P-1)}\sum_{t=a}^{b-1}z_t^{\mathsf T}z(m_S).
\label{eq:supp-s1}
\end{equation}
to the matching objective in Eq. (1) of the main article.

Two sums determine this contribution: the sum of the centered observations, which determines the direction of the centered template, and the sum of the standardized observations, which are compared with that template. Both can be obtained directly from the endpoints of the interval. Define $A_b=\sum_{t<b}c_t$ and $B_b=\sum_{t<b}z_t$. Because centering is linear,
\begin{equation}
Hm_S=\frac{A_b-A_a}{L},\qquad
z(m_S)=\sqrt{P-1}\frac{A_b-A_a}{\|A_b-A_a\|_2},
\label{eq:supp-s2}
\end{equation}
The factor $1/L$ cancels during normalization. Furthermore, the template $z(m_S)$ is common to all terms in Eq.~\eqref{eq:supp-s1}, so the sum of standardized observations can be collected into $B_b-B_a$. Substituting Eq.~\eqref{eq:supp-s2} into Eq.~\eqref{eq:supp-s1} yields
\begin{equation}
F(a,b)=\frac{(B_b-B_a)^{\mathsf T}(A_b-A_a)}
{T\sqrt{P-1}\|A_b-A_a\|_2}.
\label{eq:supp-s3}
\end{equation}
The interval score can therefore be calculated from two prefix differences, without revisiting the individual observations. The identity holds for a state mean that varies across features. It also preserves the distinction between spatial standardization of each activity pattern during GSBS and preprocessing performed on the input matrix before fitting.

The complete matching objective is the sum of the contributions of the state intervals. When a single boundary is inserted, only one interval contribution needs to be replaced by two new contributions. When two boundaries are inserted within the same state, the original contribution is replaced by three. Scores of intervals that remain unchanged can be reused. To reproduce the normalization in a local search block of length $L_b$, we multiply the globally normalized scores by $T/L_b$. The CPU implementation evaluates prefix differences in bounded tiles. PyTorch computes interval scores using batched matrix products that collect inner products between prefix vectors, and recomputes prefix differences directly for short intervals. Triton combines the prefix differences, dot product, and norm within a GPU kernel.

The reuse of interval scores also allows candidate evaluation to proceed without storing the score of every candidate. The streaming reduction visits the original blockwise candidate domain in the reference order and retains the best candidate encountered so far. The procedure follows the reference conventions for placeholder scores assigned to unevaluated candidates, the selection of the first maximum in a tie, and the handling of nonfinite scores. These conventions matter because each selected candidate determines the subsequent course of the greedy search. The matching objective identifies the best proposals, and t-distance then determines the choice between the single- and double-boundary alternatives. During refinement, boundaries are removed and reinserted sequentially, retaining their insertion labels. The complete output also preserves the reference indexing: a double insertion can associate the same accepted partition with neighboring state-count indices, and the last accepted partition may contain $K_{\max}+1$ states.

\begin{table}[!ht]
\centering\small
\caption{Principal computational costs of the reference-style and HP-GSBS calculations. $L$ denotes the length of a searched block and $K$ the current state count. The costs refer to individual operations; complete-fit runtime also depends on the candidate counts and the number of accepted updates.}\label{tab:supp-costs}
\begin{tabularx}{\textwidth}{@{}p{.30\textwidth}XX@{}}
\toprule
Operation & Reference-style calculation & HP-GSBS calculation \\
\midrule
One candidate score & Reconstruct and normalize state templates; $O(TP)$ & Full-Q lookup: $O(1)$; raw-prefix score: $O(P)$ \\
Interval precomputation & Repeated during candidate evaluation & $O(T^2P)$ time; $O(T^2)$ table storage \\
Double-candidate score storage & Store an $O(L^2)$ candidate-score array & Retain the best candidate during enumeration of the eligible pairs \\
Fixed-partition TD & Scan the $O(T^2)$ pair structure & $O(K)$ rectangle queries after obtaining endpoints \\
Feature-moment TD & Dense pairwise time matrix & $O(TP^2+KP^2+TdP)$ arithmetic, where $d=d_{\min}$ \\
Full boundary history & $O(K_{\max}T)$ & Retain the same required history \\
\bottomrule
\end{tabularx}
\end{table}

\section{T-distance from sufficient statistics}\label{sec:supp-2}
T-distance evaluates a segmentation by comparing two distributions of spatial-pattern correlations: one for pairs of time points within the same state and the other for pairs belonging to consecutive states. Let $U$ denote the row-standardized data after removal of residual row means, matching the centering in \texttt{numpy.cov}. The correlation between time points $i$ and $j$ is $r_{ij}=U_i^\top U_j/(P-1)$. Let $s(i)$ be the state label of time point $i$ and let $d=d_{\min}$ denote the temporal exclusion parameter. The two distributions are
\begin{align}
W&=\{r_{ij}:j-i\geq d,\ s(i)=s(j)\},\\
A&=\{r_{ij}:j-i\geq d,\ s(j)=s(i)+1\}.
\end{align}
Only the eligible upper-triangular pairs enter the statistic, with each pair counted once. For each distribution, its count $n$, sum $S$, and squared sum $R$ are sufficient to reconstruct the mean and unbiased variance. These quantities determine the Welch statistic \citep{welch1947}:
\begin{equation}
\mu=S/n,\qquad v=\frac{R-S^2/n}{n-1},\qquad
\mathrm{TD}=\frac{\mu_W-\mu_A}{\sqrt{v_W/n_W+v_A/n_A}}.
\label{eq:supp-s6}
\end{equation}
In GSBS, the resulting statistic serves as a criterion for comparing segmentations. Computing it from the three summary statistics of each distribution avoids repeatedly constructing the corresponding arrays of individual correlation values. We use $Q$ for the interval-template score table and $\Pi$ for the two-dimensional prefix tables used to calculate t-distance.

\subsection*{Two-dimensional prefixes}
The arrangement of eligible correlations in the time-by-time matrix is determined by the state intervals. The within-state pairs are contained in diagonal square regions, and pairs from neighboring states are contained in rectangles linking consecutive intervals. We can therefore obtain the necessary summary statistics by summing over these regions, after masking out ineligible pairs. The stable backend constructs three zero-padded two-dimensional prefix tables for the eligible-pair indicator, masked correlations, and their squares. Each prefix entry represents the sum over the rows and columns preceding its indices. Four entries then determine the sum of an arbitrary rectangle by inclusion and exclusion, as described in Section~2.3 of the main article. Combining the within-state and adjacent-state rectangle sums gives the six quantities needed in Eq.~\eqref{eq:supp-s6}.

Extracting the interval endpoints from a stored boundary vector requires $O(T)$ time. After the endpoints have been obtained, the number of required rectangles grows linearly with the number of states, so evaluating t-distance requires $O(K)$ queries. The implementation follows the reference handling of small samples. Near-zero variances or numerically ambiguous single-versus-double comparisons are reevaluated using the reference calculation.

\subsection*{Feature moments for long inputs}
The prefix-table formulation makes repeated queries inexpensive, but the number of time-point pairs increases quadratically with recording length. When there are many more time points than features, it may be more efficient to calculate the required sums of correlations directly in feature space. The relevant quantities can be derived from sums of vectors and sums of their outer products, without constructing a dense time-by-time correlation matrix.

Let $h=P-1$. For an interval $S$, we use the following summaries:
\begin{equation}\begin{gathered}
L_S=|S|,\quad u_S=\sum_{t\in S}U_t,\quad M_S=\sum_{t\in S}U_tU_t^{\mathsf T},\\
D_{1S}=\sum_{t\in S}\|U_t\|^2,\quad D_{2S}=\sum_{t\in S}\|U_t\|^4.
\end{gathered}
\end{equation}
Here, $L_S$ counts the observations, $u_S$ sums their feature vectors, and $M_S$ sums their outer products. The scalar summaries $D_{1S}$ and $D_{2S}$ account for self-pairs when calculating correlations between distinct time points. Before excluding short-lag pairs, the number of unordered within-state pairs is
\begin{equation}
n_S=L_S(L_S-1)/2
\end{equation}
To obtain the sum of correlations, expand the squared norm of the vector sum:
\[
\left\|\sum_{t\in S}U_t\right\|_2^2=\sum_{t\in S}\|U_t\|_2^2+2\sum_{i<j,\ i,j\in S}U_i^\top U_j.
\]
The first term on the right contains the self-pairs. The factor of two in the second term reflects the two orders in which each distinct pair occurs in the full expansion. Subtracting the self-pair contribution and dividing by $2h$ gives the required sum of correlations:
\begin{equation}
S_S=(\|u_S\|^2-D_{1S})/(2h)
\end{equation}
For the squared sum of correlations, the analogous identity uses the outer products. Since the Frobenius inner product between $U_iU_i^\top$ and $U_jU_j^\top$ equals $(U_i^\top U_j)^2$, we have
\[
\left\|\sum_{t\in S}U_tU_t^\top\right\|_F^2=\sum_{t\in S}\|U_t\|_2^4+2\sum_{i<j,\ i,j\in S}(U_i^\top U_j)^2.
\]
Again, subtracting the self-pair contribution and applying the correlation normalization yields
\begin{equation}
R_S=(\|M_S\|_F^2-D_{2S})/(2h^2).
\end{equation}
Thus, the count, sum, and squared sum of the within-state correlations can all be expressed through the feature-space summaries. For two distinct consecutive intervals $S$ and $R$, each time point in one interval forms a cross-interval pair with each time point in the other. Their correlation sum is given by the inner product of the two vector sums; the sum of squared correlations is given by the Frobenius inner product of the two outer-product sums. The corresponding statistics are
\begin{equation}
n_{SR}=L_SL_R,\qquad S_{SR}=u_S^{\mathsf T}u_R/h,\qquad
R_{SR}=\langle M_S,M_R\rangle_F/h^2.
\end{equation}
Equations (S8)--(S11) initially include pairs with short positive temporal separations. To implement the temporal exclusion rule, we subtract the count, sum, and squared sum of pairs with $0<j-i<d$ from the corresponding distribution. The implementation retains state summaries and bounded caches of short-lag contributions, allowing the working storage for the correlation calculation to remain in feature space. Candidate scoring uses temporal prefixes of the raw-input and standardized-input representations, with working arrays of at most $q\times P$ elements for chunks of up to $q$ candidates. The complete boundary history still requires $O(K_{\max}T)$ storage.

The moment formulas involve subtracting quantities that can be large relative to their difference. When correlations are nearly constant, rounding errors in these subtractions can affect the estimated variance. In such cases, the low-memory calculation evaluates pairwise correlations in bounded blocks using two passes to compute their statistics. Unresolved near-degenerate cases trigger a numerical error. The feature-moment formulation is available as an experimental option for long recordings with relatively few features. The FP64 Full-Q and Prefix2D formulation is used for the statewise analyses reported in the main article.

\section{Execution interfaces and numerical scope}\label{sec:supp-3}
The standard execution path uses FP64 arithmetic, Full-Q interval scores, streaming candidate evaluation, and Prefix2D t-distance. The CPU, PyTorch CUDA, and Triton backends construct statistics for a common search controller, which performs boundary selection and refinement in the reference order. The interface also accepts an optional independent matrix $Y$ for calculating t-distance. This allows candidates and state patterns to be determined from $X$ while the segmentation is evaluated using correlations calculated from $Y$.

Small differences in floating-point arithmetic can influence the ordering of nearly equal candidate scores. In these cases, the implementation reevaluates candidates using the original expressions. Together with the variance calculation described in Section~\ref{sec:supp-2}, this procedure supports consistent boundary decisions across the evaluated workloads. Input values must be finite, and each time point must have nonzero variance across features so that spatial Pearson correlations are defined. Numerically ill-conditioned inputs are handled using the reference expressions.

Long recordings and extensive candidate searches can require substantial computation and memory. Users can set limits on candidate evaluation, reevaluation with the original expressions, working memory, and blockwise pair-statistic calculations. The fit reports the exceeded resource limit and stops with the requested scientific configuration preserved.

\begin{table}[!ht]
\centering\small
\caption{Standard, experimental, and research-stage execution modes. The standard FP64 configuration supports the statewise analyses reported in the main article.}\label{tab:modes}
\begin{tabularx}{\textwidth}{@{}p{.27\textwidth}p{.17\textwidth}>{\raggedright\arraybackslash}X@{}}
\toprule
Mode & Status & Role \\
\midrule
FP64 Full-Q + Prefix2D, three backends & Stable & Standard numerical path with the shared sequential search and refinement procedure. \\
Raw-prefix + feature moments & Experimental & CPU mode for long recordings, with blockwise variance calculation in numerically sensitive cases. \\
On-GPU Triton candidate reduction & Experimental & Selects the highest-scoring candidate directly on the GPU for selected large candidate sets. \\
FP32 & Experimental & Single-precision GPU computation for studying speed and precision; t-distance values differ from FP64. \\
Incremental TD; no-Q GPU prototypes & Research & Incremental updates of partition statistics and component-level GPU experiments with interval scoring from prefixes. \\
\bottomrule
\end{tabularx}
\end{table}

The saved result contains the complete boundary history, t-distance curve, selected states, state patterns, and boundary strengths. This allows the same fitted recording to be examined at different state counts. Collections of recordings are processed independently: GPU batches group compatible shapes and resource settings, while CPU workers handle separate recordings. When a batch analysis resumes, completed results can be reused if the input data, analysis and execution settings, and software version match the requested fit. Outputs from earlier versions remain readable; rerunning a recording with a new implementation uses a new output directory.

\subsection*{Example analysis workflow}
The executable example \path{scripts/research_workflow.py} illustrates a complete single-recording workflow. It generates a small prepared input, fits it with the CPU backend, retrieves boundaries and t-distance, saves and reloads a versioned result, and then processes two recordings independently. The same result interface is used by the CPU example in the main article. On a compatible CUDA installation, the backend can be changed to \texttt{torch-cuda} or \texttt{triton-cuda}. For CPU multiprocessing, the number of workers can be configured through \texttt{GSBSConfig(n\_jobs=...)}. On platforms using the spawn start method, multiprocessing calls should be placed inside a \texttt{\_\_main\_\_} guard.

When the analysis involves a collection of prepared recordings, the released command-line interface accepts a task manifest. For example:
\begin{lstlisting}[language=bash]
python -m hpgsbs fit-batch tasks.json --backend cpu-fast \
    --workers 4 --output outputs
\end{lstlisting}
The task list specifies the input paths and analysis settings. Each result file becomes available once it has been written in full. Running the same command resumes an interrupted analysis by reusing completed results whose input data, analysis and execution settings, and software version match the requested fit. The software documentation provides the task-list format and further examples \citep{hpgsbs2026}.

\section{Experimental protocols and data provenance}\label{sec:supp-4}
\subsection*{Controlled paired benchmark}
To compare the effects of recording length, feature dimension, and search depth, we used a one-factor-at-a-time design with the same central configuration, $T=512$, $P=32$, and $K_{\max}=20$, in each of the three series. Seven unique configurations were evaluated using seeds 20261017 and 20261018. For each seed, Gaussian state means were generated as an $8\times64$ matrix, followed by a $T\times64$ Gaussian noise matrix scaled by 0.8. The eight contiguous states were defined using rounded, equally spaced endpoints. For a given feature count, we retained the first $P$ features. Consequently, the feature-count comparisons use nested subsets of features, and the state-count comparisons reuse identical input matrices. Across configurations, the 14 configuration--seed cases contain ten distinct input matrices. Reusing the same seeds also couples observations across conditions, facilitating controlled comparisons.

For each official observation, the timed interval includes object construction and the call to \texttt{GSBS.fit(showProgressBar=False)}. Imports and numerical initialization were performed using a separate short input before timing. Each HP-GSBS observation consists of one untimed warm-up followed by three timed complete fits through the released public API. The order of the backends was rotated across configurations and seeds, with GPU execution synchronized immediately before and after timing. Validation and serialization were performed outside the timed interval. The complete benchmark therefore comprises 14 timed official fits, 126 timed HP-GSBS fits, and 42 HP-GSBS warm-up fits. Across the seven configurations, all 168 optimized outputs satisfied the numerical agreement criteria.

We calculate the speedup separately for each input so that the official and optimized implementations are compared on the same recording. Let $t_{i,b,r}$ be the runtime for input $i$, optimized backend $b$, and repetition $r$, and let $t_{i,\mathrm{official}}$ be the official runtime for that input. We first take the median of the three optimized runtimes and then calculate the corresponding same-input speedup:
\begin{equation}
\widetilde{t}_{i,b}=\operatorname{median}_r t_{i,b,r},\qquad
G_{i,b}=t_{i,\mathrm{official}}/\widetilde{t}_{i,b}.
\end{equation}
For each configuration, the summarized runtimes are the medians of the two input-level observations, and the summarized speedups are the medians of the two paired ratios. The whiskers in the main figure represent the observed range across the two seeds. Each official input-level observation comes from one timed fit, while each optimized observation summarizes three. Tables~S5--S7 retain this distinction, and the data bundle contains all 140 timed observations.

\subsection*{Mechanism illustration and data provenance}
Figure~1 uses a short excerpt of the public Sherlock movie-viewing dataset to illustrate how the same neural activity matrix supplies both state-template scores and correlation statistics \citep{baldassano2020sherlock,chen2017shared}. The input is the mean angular-gyrus response across 16 participants, restricted to the first 120 TRs (TR = 1.5 s). Fitting uses all 4,230 voxels, while the activity heatmap displays a subset of 64 voxels. The interval scores and correlation matrices shown in the figure are derived from this input, and smaller subsets are used to illustrate rectangle queries. The full symmetric correlation matrix is displayed to make the state structure visible. The actual statistics use the temporal mask and eligible upper-triangular pairs, so each eligible pair is counted once. The interval-template table $Q$ stores the matching scores of candidate intervals, while the correlation prefixes store cumulative summaries of pairwise similarities.

The saved illustration was generated with the CPU backend of HP-GSBS 0.2.0 in FP64, using $K_{\max}=30$, $d_{\min}=1$, \texttt{finetune=1}, and a block size of 50. The associated numerical checks compare its output with \texttt{statesegmentation} 0.0.6. The figure illustrates the architecture shared by the three backends evaluated in the synthetic and EEG experiments. The figure materials include the numerical arrays, search history, source attribution, and comparison with GSBS.

\subsection*{Illustrative signal and EEG provenance}
The synthetic example in Fig.~2 provides a recording for which the generating boundaries are known. Using seed 19, we generate four Gaussian mean patterns with 16 features, repeat each pattern for 20 samples, and add Gaussian noise with standard deviation 0.05. The resulting 80-sample recording has generating boundaries at samples 20, 40, and 60. The official implementation and all three HP-GSBS backends select these boundaries; Fig.~2 presents the selected state count and the numerical differences in t-distance. Supplementary Table~\ref{tab:recovery} reports examples with other seeds and noise levels to illustrate how recovery of the generating boundaries changes with noise, even when the implementations make the same segmentation decisions.

The EEG evaluation uses four fixed, archived analysis inputs derived from the Brain, Body, and Behavior Dataset (BBBD; \citealp{bbbd2026}), created by Jens Madsen, Nikhil Kuppa, and Lucas C. Parra and distributed under CC BY 4.0. Table~\ref{tab:eeginputs} lists the recording identifiers and retained dimensions. Preprocessing included cropping around stimulus events, masking artifacts, averaging consecutive nonoverlapping groups of 25 samples to reduce the sampling rate from 128 to 5.12 Hz, and standardizing each channel within a recording using the sample standard deviation of valid time bins. The stored float32 inputs are converted to FP64 for fitting. All four recordings use $d_{\min}=5$, statewise detection, \texttt{finetune=1}, ordered sequential refinement, and a block size of 50, with recording-specific $K_{\max}$ values. The archived EEG arrays, together with their source recording identifiers and preprocessing history, provide the inputs for reproducing the reported segmentations.

\begin{table}[!ht]
\centering\footnotesize\setlength{\tabcolsep}{4pt}
\caption{The four archived BBBD analysis inputs. Recording numbers in the article are one-based, whereas file identifiers are zero-based. Source recording identifiers and experiment information follow the archived acquisition metadata (upstream dataset release version unrecorded). Recording duration is $T/5.12$ s, and the time maps record the bin centers. For Recording 3, the selected $K_{\max}+1$ index follows the reference double-insertion convention described in Section~\ref{sec:supp-1}.}\label{tab:eeginputs}
\begin{tabular}{@{}rllrlrr@{}}\toprule
Rec. & Recording identifier & Exp. & $T\times P$ & Channels & $K_{\max}$ & Selected $K$ \\\midrule
1 & \texttt{sub-01\_ses-01\_stim01} & Unspecified & $1336\times64$ & All 64 & 131 & 126 \\
2 & \texttt{sub-02\_ses-01\_stim02} & 3 & $1857\times32$ & Frontal 32 & 121 & 120 \\
3 & \texttt{sub-03\_ses-02\_stim03} & 3 & $1754\times64$ & All 64 & 172 & 173 \\
4 & \texttt{sub-04\_ses-02\_stim06} & 3 & $1374\times64$ & All 64 & 135 & 131 \\
\bottomrule\end{tabular}
\end{table}

\begin{table}[!ht]
\centering\footnotesize
\caption{Mapping between the archived EEG inputs and the reference outputs used for numerical comparison. Paths are relative to the accompanying source-package root and fixed public EEG snapshot. Each input contains \texttt{x} and \texttt{times}. Recording 1 uses the input from the Fig.~3 source files. Backend paths, algorithm settings, channel names, and hashes are listed in \texttt{data/eeg/provenance.json}.}\label{tab:eegfiles}
\begin{tabularx}{\textwidth}{@{}r>{\raggedright\arraybackslash}X>{\raggedright\arraybackslash}X@{}}\toprule
Rec. & Input file & Comparison reference \\\midrule
1 & \path{figure_sources/figure3/data/sub-01_ses-01_stim01__all64__5.12hz__d5__k131.npz} & \path{figure_sources/figure3/data/0_full_reference.npz}(complete output from the official GSBS implementation) \\
2 & \path{data/eeg/inputs/recording_1.npz} & \path{data/eeg/outputs/recording_1_full_cpu.npz}(CPU output previously validated against official GSBS) \\
3 & \path{data/eeg/inputs/recording_2.npz} & \path{data/eeg/outputs/recording_2_full_cpu.npz}(CPU output previously validated against official GSBS) \\
4 & \path{data/eeg/inputs/recording_3.npz} & \path{data/eeg/outputs/recording_3_full_cpu.npz}(CPU output previously validated against official GSBS) \\
\bottomrule\end{tabularx}
\end{table}

Table~\ref{tab:eegfiles} links each EEG analysis input to its corresponding reference output. The backend outputs use the suffixes \path{_full_cpu.npz}, \path{_torch-cuda_float64.npz}, and \path{_triton-cuda_float64.npz}. For Recording 1, the shared filename stem is \path{figure_sources/figure3/data/0}; for Recordings 2--4, it is \path{data/eeg/outputs/recording_1} through \path{recording_3}. File checksums and processed-array hashes, together with their definitions, are documented in \path{data/eeg/provenance.json}.

Table~3 of the main article summarizes the four full-record comparisons from the archived EEG snapshot identified by commit \texttt{f90797909882a067578f7a50c6814de46d333d98}. For Recording 1, the reference is the complete output of the official GSBS implementation. For Recordings 2--4, saved CPU outputs previously verified against the official implementation provide the reference for evaluating PyTorch CUDA and Triton. The reproducibility archive also provides comparisons of all three HP-GSBS 0.2.0 backends with the saved reference outputs. Both sets of comparisons examine the complete boundary history, selected state count, t-distance, state patterns, and boundary strengths using the numerical criteria below.

Figure~3 illustrates the first EEG recording using the fixed $1{,}336\times64$ input and four complete saved outputs. The segmentation is fitted to the entire recording; the local view is then selected from this full-recording result using fixed sample indices 617--718, corresponding to 120.51--140.43 s. The input has already been standardized per channel. For visualization, heatmap colors are clipped at $\pm3$ without altering the analysis arrays. Spatial correlations are calculated using all 64 channels and displayed as a symmetric time-by-time matrix, with selected state blocks clipped at the edges of the displayed window. Both the full-record state sequence and the local boundary positions are taken directly from the saved selected partition.

The plot displays the evaluated state-count indices $K=2$--131. All four outputs select 126 states and agree exactly in the complete boundary histories and state patterns. The maximum absolute differences in t-distance from the official output are $2.43\times10^{-13}$ for CPU and $9.59\times10^{-14}$ for each GPU backend. Thus, the local window illustrates the segmentation and numerical agreement obtained from the whole recording.

\subsection*{Numerical criteria}
We compare boundary arrays and selected state counts by exact equality. For t-distance, state patterns, and boundary strengths, we use absolute and relative tolerances of $10^{-8}$, requiring matching array dimensions and corresponding nonfinite entries. Across the controlled benchmark, the maximum absolute difference in t-distance is $1.88\times10^{-11}$, and the state patterns and boundary strengths are identical. For the illustrative signal, we also compare complete search and refinement traces, including individual boundary-refinement moves.

\subsection*{Environment}
The controlled benchmark ran on an AMD Ryzen 7 7840H CPU and NVIDIA RTX 4050 Laptop GPU (6 GB) under Ubuntu 22.04/WSL2, with Python 3.10.12, NumPy 2.2.6, SciPy 1.15.3, Numba 0.68.0, PyTorch 2.11.0+cu128, Triton 3.6.0, and CUDA runtime 12.8. BLAS and Numba thread counts and PyTorch intra- and inter-operation thread counts were set to one; TF32 was disabled. The complementary optimization, memory, and throughput measurements were collected during software release benchmarking on the same laptop family and are reported in Section~\ref{sec:supp-6}.

\FloatBarrier
\section{Complete controlled-benchmark tables}\label{sec:supp-5}
Tables~S5--S7 present the same controlled benchmark at three levels of detail. Table~\ref{tab:runtime} summarizes the complete-fit runtimes for each configuration, Table~\ref{tab:speedups} summarizes the paired speedups relative to the official implementation, and Table~\ref{tab:perinput} provides the individual input-level timing observations underlying these summaries. The central configuration is reused in the recording-length, feature-count, and search-depth comparisons, as described in Section~\ref{sec:supp-4}.
\begin{table}[!ht]
\centering\footnotesize\setlength{\tabcolsep}{3pt}
\caption{Complete-fit runtimes in seconds, reported as median [minimum, maximum] across the two input seeds. Each official input-level observation is a single timed fit; each HP-GSBS input-level observation is the median of three timed fits after warm-up.}\label{tab:runtime}
\begin{tabular}{@{}rrrllll@{}}\toprule
$T$ & $P$ & $K_{\max}$ & Official & CPU & PyTorch & Triton \\\midrule
256 & 32 & 20 & 11.234 [10.144, 12.324] & 0.073 [0.073, 0.073] & 0.068 [0.065, 0.070] & 0.064 [0.063, 0.064] \\
512 & 32 & 20 & 64.667 [58.876, 70.458] & 0.142 [0.138, 0.146] & 0.081 [0.075, 0.087] & 0.099 [0.086, 0.111] \\
768 & 32 & 20 & 189.610 [169.973, 209.247] & 0.251 [0.226, 0.277] & 0.110 [0.105, 0.116] & 0.131 [0.124, 0.138] \\
512 & 8 & 20 & 53.323 [52.454, 54.192] & 0.122 [0.120, 0.123] & 0.081 [0.077, 0.085] & 0.075 [0.073, 0.076] \\
512 & 64 & 20 & 79.365 [79.157, 79.573] & 0.153 [0.149, 0.156] & 0.077 [0.073, 0.082] & 0.100 [0.094, 0.106] \\
512 & 32 & 10 & 55.603 [50.179, 61.028] & 0.118 [0.117, 0.120] & 0.065 [0.063, 0.068] & 0.076 [0.069, 0.082] \\
512 & 32 & 40 & 75.652 [69.669, 81.635] & 0.165 [0.160, 0.170] & 0.111 [0.106, 0.116] & 0.126 [0.126, 0.126] \\

\bottomrule\end{tabular}
\end{table}

\begin{table}[!ht]
\centering\small\setlength{\tabcolsep}{4pt}
\caption{Speedup factors relative to the official implementation, calculated separately on each input and summarized as the median [minimum, maximum] of the two paired ratios. Ratios are calculated from the unrounded timing observations before summarization.}\label{tab:speedups}
\begin{tabular}{@{}rrrlll@{}}\toprule
$T$ & $P$ & $K_{\max}$ & CPU & PyTorch & Triton \\\midrule
256 & 32 & 20 & 154.0 [139.5, 168.5] & 166.9 [144.1, 189.7] & 177.0 [158.6, 195.4] \\
512 & 32 & 20 & 454.1 [425.1, 483.2] & 794.8 [782.0, 807.6] & 675.0 [528.2, 821.9] \\
768 & 32 & 20 & 770.1 [613.5, 926.6] & 1729.7 [1470.9, 1988.4] & 1458.3 [1235.0, 1681.7] \\
512 & 8 & 20 & 438.5 [427.1, 450.0] & 662.2 [616.9, 707.4] & 715.7 [690.1, 741.2] \\
512 & 64 & 20 & 520.5 [508.8, 532.2] & 1027.7 [967.7, 1087.7] & 797.3 [751.1, 843.5] \\
512 & 32 & 10 & 469.9 [429.9, 509.8] & 854.8 [740.3, 969.4] & 735.1 [723.7, 746.6] \\
512 & 32 & 40 & 457.6 [434.4, 480.7] & 683.6 [599.8, 767.4] & 599.3 [551.7, 646.9] \\

\bottomrule\end{tabular}
\end{table}

\begin{table}[!ht]
\centering\small\setlength{\tabcolsep}{4pt}
\caption{Input-level complete-fit runtimes (seconds) underlying Tables~S5 and S6. The official column contains one timed fit per input; each HP-GSBS column contains the median of three timed fits for the corresponding backend.}\label{tab:perinput}
\begin{tabular}{@{}rrrr rrrr@{}}\toprule
$T$ & $P$ & $K_{\max}$ & Seed & Official & CPU & PyTorch & Triton \\\midrule
256 & 32 & 20 & 20261017 & 12.3239 & 0.0731 & 0.0650 & 0.0631 \\
256 & 32 & 20 & 20261018 & 10.1440 & 0.0727 & 0.0704 & 0.0639 \\
512 & 32 & 20 & 20261017 & 70.4578 & 0.1458 & 0.0872 & 0.0857 \\
512 & 32 & 20 & 20261018 & 58.8762 & 0.1385 & 0.0753 & 0.1115 \\
768 & 32 & 20 & 20261017 & 209.2472 & 0.2258 & 0.1052 & 0.1244 \\
768 & 32 & 20 & 20261018 & 169.9726 & 0.2770 & 0.1156 & 0.1376 \\
512 & 8 & 20 & 20261017 & 52.4543 & 0.1228 & 0.0850 & 0.0760 \\
512 & 8 & 20 & 20261018 & 54.1916 & 0.1204 & 0.0766 & 0.0731 \\
512 & 64 & 20 & 20261017 & 79.5733 & 0.1564 & 0.0822 & 0.1059 \\
512 & 64 & 20 & 20261018 & 79.1572 & 0.1487 & 0.0728 & 0.0938 \\
512 & 32 & 10 & 20261017 & 61.0280 & 0.1197 & 0.0630 & 0.0817 \\
512 & 32 & 10 & 20261018 & 50.1787 & 0.1167 & 0.0678 & 0.0693 \\
512 & 32 & 40 & 20261017 & 81.6352 & 0.1698 & 0.1064 & 0.1262 \\
512 & 32 & 40 & 20261018 & 69.6693 & 0.1604 & 0.1162 & 0.1263 \\

\bottomrule\end{tabular}
\end{table}

\FloatBarrier
\section{Complementary performance and recovery results}\label{sec:supp-6}
\subsection*{Extended benchmarks}
To examine performance on larger inputs, Table~\ref{tab:extended} presents the extended HP-GSBS 0.2.0 benchmarks \citep{hpgsbsreleasebench2026}. These include synthetic recordings of up to 2,100 time points and a $1{,}336\times64$ EEG recording. The official and HP-GSBS fits were timed in separate sessions on the same machine. Synthetic runtimes are summarized as the median of three timed fits after warm-up; the EEG comparison uses one complete fit per implementation. Timings include computation and required data transfers, with result comparison and file output performed afterward.

\begin{table}[!ht]
\centering\footnotesize\setlength{\tabcolsep}{3pt}
\caption{Complete-fit runtimes for extended synthetic workloads and an EEG recording (seconds), with speedups relative to the official implementation in parentheses. The implementations were timed in separate sessions on the same machine. Speedups are calculated from the original timing observations before rounding.}\label{tab:extended}
\begin{tabular}{@{}lrrrlll@{}}\toprule
Workload & $T\times P$ & $K_{\max}$ & Official & CPU & PyTorch & Triton \\\midrule
Small & $450\times16$ & 20 & 48.0 & 0.103 (467x) & 0.075 (636x) & 0.057 (847x) \\
Medium & $900\times32$ & 30 & 349.9 & 0.317 (1,105x) & 0.098 (3,556x) & 0.098 (3,577x) \\
Large & $1500\times64$ & 40 & 2213.5 & 0.868 (2,550x) & 0.172 (12,903x) & 0.213 (10,377x) \\
XL & $2100\times64$ & 50 & 5011.3 & 1.952 (2,567x) & 0.363 (13,816x) & 0.456 (10,988x) \\
EEG & $1336\times64$ & 131 & 6115.4 & 1.610 (3,798x) & 0.853 (7,165x) & 0.838 (7,299x) \\

\bottomrule\end{tabular}
\end{table}
For the largest synthetic input, the official GSBS fit took about 5,011 s, while the PyTorch implementation completed the fit in 0.363 s, corresponding to an approximately 13,816-fold speedup. Together with the controlled comparisons in Section~\ref{sec:supp-5}, these measurements illustrate the computational benefit of reusing interval and correlation statistics as recording length increases.

\subsection*{Ablation and memory}
The CPU ablation examines the costs that remain after interval-score precomputation has been introduced. It uses one fixed EEG input ($1{,}336\times64$, $K_{\max}=131$, $d_{\min}=5$), with statewise detection, sequential refinement, and a block size of 50. Starting from the Full-Q configuration that stores candidate scores, we successively introduce streaming candidate reduction, local-window refinement, and reuse of the best candidates in unchanged search blocks. Table~\ref{tab:ablation} shows how each successive optimization reduces the complete-fit runtime. All four configurations produce identical segmentations.

\begin{table}[!ht]
\centering\small
\caption{Stepwise CPU ablation starting from a Full-Q configuration that stores candidate scores. Complete-fit runtimes (seconds) are summarized over three timed fits after one warm-up. All four configurations produce identical segmentations.}\label{tab:ablation}
\begin{tabular}{@{}lrrr@{}}\toprule
Configuration & Median & Minimum & Maximum \\\midrule
Stored candidate scores (Full-Q) & 3.3638 & 3.2950 & 3.3870 \\
Streaming candidate evaluation & 1.8035 & 1.7705 & 1.9462 \\
Streaming + local refinement window & 1.6014 & 1.5863 & 1.6605 \\
Streaming + local window + block cache & 1.4288 & 1.4275 & 1.4291 \\
\bottomrule\end{tabular}
\end{table}

We separately compare the memory and runtime costs of the precomputed-table and low-memory calculations for a $5{,}000\times8$ input. Both configurations use single-boundary search, no refinement, a block size of zero, and $K_{\max}=16$. The Full-Q calculation with stored candidate scores takes 9.754 s and reaches a peak process RSS of 1,598.0 MiB; the low-memory calculation using raw prefixes and feature moments takes 0.110 s and reaches 190.6 MiB. RSS measures the total resident memory of the process, including runtime state, and was sampled every 5 ms. The observed timing ranges are 9.384--10.432 s and 0.105--0.113 s, respectively.

\begin{table}[!ht]
\centering\small
\caption{Low-memory measurements for increasing recording lengths and feature counts, using single-boundary search, no refinement, block size zero, and $K_{\max}=16$. Each median summarizes three timed fits. The final column gives the largest absolute t-distance difference for the selected and final partitions relative to an independent blockwise calculation of pairwise correlations.}\label{tab:low-memory}
\begin{tabular}{@{}rrrrr@{}}\toprule
$T\times P$ & Median (s) & Range (s) & Peak RSS (MiB) & Max. TD difference \\\midrule
$5000\times8$ & 0.1077 & 0.1074--0.1079 & 191.1 & $1.78\times10^{-15}$ \\
$10000\times16$ & 0.1843 & 0.1724--0.1850 & 204.5 & $8.88\times10^{-16}$ \\
$20000\times32$ & 0.3745 & 0.3652--0.4073 & 247.5 & $8.88\times10^{-16}$ \\
$50000\times64$ & 1.4439 & 1.4177--1.4842 & 469.5 & $1.15\times10^{-14}$ \\
\bottomrule\end{tabular}
\end{table}
Table~\ref{tab:low-memory} extends the low-memory experiment to progressively larger values of both recording length and feature count, using the same single-boundary configuration with refinement disabled. To evaluate t-distance independently while limiting memory use, we calculate the pairwise correlations in blocks for the selected and final partitions. For the $50{,}000\times64$ input, these partitions contain 15 and 16 states, respectively. A statewise search additionally enumerates pairs of candidate boundaries, whose number grows quadratically within each long interval. This additional computation can exceed the resource limits described in Section~\ref{sec:supp-3}.

\subsection*{Batch throughput}
While single-fit runtime measures the time needed to process one recording, throughput measures how many recordings can be analyzed within a given time. We evaluate throughput using two kinds of workloads. The short tasks are independent synthetic recordings of size $450\times16$, with $K_{\max}=20$ and $d_{\min}=1$. The high-state-count tasks are repeated copies of one $1{,}336\times64$ EEG input, with $K_{\max}=131$ and $d_{\min}=5$. Each configuration has one warm-up batch followed by three timed batches. Timing includes pool construction, serialization where applicable, fitting, and synchronization, with validation performed afterward. Each batch is subject to the configured memory budget. We calculate throughput as the number of completed fits per minute in each timed batch. CPU batches contain 32 short tasks or eight high-state-count tasks. Table~\ref{tab:throughput} lists the GPU batch sizes, including the larger batches evaluated in this experiment.

\begin{longtable}{@{}llrrl@{}}
\caption{Batch throughput, measured as completed fits per minute and reported as median [minimum, maximum] across three timed batches. Short tasks use independent synthetic inputs; high-state-count tasks repeat the same EEG input. CPU worker counts and GPU batch sizes are reported separately.}\label{tab:throughput}\\
\toprule Workload & Backend & Workers & Batch & Fits/minute \\\midrule\endfirsthead
\toprule Workload & Backend & Workers & Batch & Fits/minute \\\midrule\endhead
\bottomrule\endfoot
Small & CPU & 1 & 32 & 274.3 [269.4, 282.9] \\
Small & CPU & 2 & 32 & 839.6 [815.0, 843.6] \\
Small & CPU & 4 & 32 & 1362.2 [1328.0, 1419.7] \\
Small & CPU & 8 & 32 & 1766.8 [1638.0, 1781.1] \\
Small & PyTorch & 1 & 1 & 646.7 [641.3, 730.3] \\
Small & PyTorch & 1 & 8 & 1418.4 [1383.6, 1425.6] \\
Small & PyTorch & 1 & 32 & 1459.9 [1378.7, 1538.1] \\
Small & PyTorch & 1 & 64 & 1399.3 [1398.8, 1484.2] \\
Small & Triton & 1 & 1 & 658.6 [652.8, 660.1] \\
Small & Triton & 1 & 8 & 1451.5 [933.0, 1654.5] \\
Small & Triton & 1 & 32 & 1529.9 [1384.9, 1598.3] \\
Small & Triton & 1 & 64 & 1340.3 [1313.8, 1358.1] \\
High-K & CPU & 1 & 8 & 24.4 [24.0, 24.9] \\
High-K & CPU & 2 & 8 & 74.3 [74.1, 74.7] \\
High-K & CPU & 4 & 8 & 125.7 [124.5, 129.4] \\
High-K & CPU & 8 & 8 & 154.9 [149.0, 161.9] \\
High-K & PyTorch & 1 & 1 & 42.9 [42.0, 47.9] \\
High-K & PyTorch & 1 & 4 & 40.5 [38.8, 40.7] \\
High-K & PyTorch & 1 & 8 & 41.2 [40.4, 41.5] \\
High-K & Triton & 1 & 1 & 41.9 [40.5, 45.6] \\
High-K & Triton & 1 & 4 & 44.6 [44.2, 45.1] \\
High-K & Triton & 1 & 8 & 43.1 [42.2, 43.3] \\

\end{longtable}

\FloatBarrier
\subsection*{Generating-boundary recovery}
Table~\ref{tab:recovery} illustrates how the recovery of known boundaries changes across two noise levels in simulated four-state recordings. The generating boundaries are at samples 20, 40, and 60. Precision is the fraction of selected boundaries that match these locations exactly, and recall is the fraction of generating boundaries recovered. All four implementations make identical segmentation decisions for each example.
\begin{table}[!ht]
\centering\small
\caption{Recovery of known state boundaries in four-state simulations at two noise levels. The generating boundaries are at samples 20, 40, and 60. Precision and recall use exact matches to these indices. The official implementation and all HP-GSBS backends agree for each example.}\label{tab:recovery}
\begin{tabular}{@{}rrr lrr@{}}\toprule
Seed & Noise SD & Selected $K$ & Selected boundaries & Precision & Recall \\\midrule
12 & 0.05 & 3 & 20, 40 & 1.00 & 0.67 \\
12 & 2.00 & 2 & 42 & 0.00 & 0.00 \\
19 & 0.05 & 4 & 20, 40, 60 & 1.00 & 1.00 \\
19 & 2.00 & 6 & 3, 7, 20, 41, 58 & 0.20 & 0.33 \\
\bottomrule\end{tabular}
\end{table}
The examples illustrate how noise and state-count selection influence boundary recovery. Figure~2 of the main article uses the seed-19 example with noise standard deviation 0.05, for which all three generating boundaries are recovered and numerical agreement can be examined throughout the search.

\section{Reproducibility package and data access}\label{sec:supp-7}
The software release and the reproducibility materials serve distinct purposes: the software provides the implementation used for fitting, and the archived analysis materials provide the numerical evidence and scripts associated with this article. HP-GSBS 0.2.0 is available under the MIT license at \url{https://doi.org/10.5281/zenodo.23422148} \citep{hpgsbs2026}. The original benchmark archive, version 1.0.1, is available at \url{https://doi.org/10.5281/zenodo.23457342} \citep{hpgsbsrepro2026}. The expanded reproducibility archive, version 1.1.0, is available at \url{https://doi.org/10.5281/zenodo.23573202} \citep{hpgsbsrepro1102026} and corresponds to Git tag \texttt{v1.1.0} (commit \texttt{126043d11050368dbbcffae8f512eda65e6c4648}). It includes the four-record EEG collection, the complete Sherlock application, all six main figures, benchmark observations, provenance records, licenses, and verification scripts. The all-versions DOI is \url{https://doi.org/10.5281/zenodo.23457341}.

A reader can first verify the archived numerical materials and regenerate the figures without rerunning all fits. From a checkout of the public reproducibility repository at tag \texttt{v1.1.0}, use:
\begin{lstlisting}[language=bash]
python scripts/verify_manifest.py
python scripts/verify_data.py
python scripts/make_figures.py
\end{lstlisting}
The public acquisition and fitting scripts \path{scripts/download_sherlock.py} and \path{scripts/run_sherlock.py} also allow the fMRI application to be rerun, as described in Section~\ref{sec:supp-8}.

The public reproducibility archive provides the numerical inputs, saved outputs, and plotting scripts used to generate all six main figures. For the EEG analyses, the verification workflow checks all four archived inputs, their preprocessing and fit configurations, complete boundary histories, selected states, t-distance, state patterns, and boundary strengths. Boundary comparisons use exact equality, and floating-point comparisons use absolute and relative tolerances of $10^{-8}$. The fMRI verification workflow checks all 17 segmentations, the figure cache, and the observed boundary alignment. It can also recompute the circular-shift distribution using the recorded random seed (Section~\ref{sec:supp-8}).

To rerun the EEG analysis with the fixed software release and save the new results separately, use:
\begin{lstlisting}[language=bash]
python scripts/run_eeg_all.py --backend cpu-fast \
    --output new_eeg_cpu
\end{lstlisting}
The \texttt{--recordings} option uses zero-based identifiers 0--3. On a compatible installation, \texttt{torch-cuda} and \texttt{triton-cuda} select the corresponding FP64 GPU backends. The script compares the newly saved optimized outputs with the archived reference outputs listed in Table~\ref{tab:eegfiles}. The new results and timing measurements are saved in the separate output directory specified in the command.

All data paths are relative to the public reproducibility repository root. The controlled benchmark can be inspected through \path{data/paired_timed_runs.csv}, which contains all 140 timed observations, and \path{data/paired_inputs.csv}, which records the shapes, seeds, and input hashes. The per-input, runtime, and speedup CSV files provide successive summaries of these observations. For the illustrative simulation, \path{data/synthetic_figure_data.npz} contains the signal and backend outputs. CSV files under \path{data/} provide the ablation, memory, throughput, and EEG comparison results. For the fMRI analysis, \path{data/fmri/figure6_data.npz} contains the Sherlock display matrix, state-count curves, selected boundaries, and null scores used to construct Fig.~6.

BBBD and Sherlock data retain attribution to their original creators \citep{bbbd2026,baldassano2020sherlock,chen2017shared}. Processing histories and available recording identifiers accompany the derived inputs. Code is MIT licensed; the distributed data, derived figures, and documentation retain CC BY 4.0 attribution. The archive thus preserves the numerical observations and materials needed to inspect the analyses reported here.

\section{Naturalistic fMRI protocol and complete results}\label{sec:supp-8}
\subsection*{Input identity and analysis configuration}
The fMRI application uses the public angular-gyrus array \path{Sherlock_AG_movie.npy}, distributed in version 1 of \citet{baldassano2020sherlock} (\url{https://doi.org/10.6084/m9.figshare.12436955.v1}) and derived from \citet{chen2017shared}. The source file is available at \url{https://ndownloader.figshare.com/files/22927253}. It contains 256,101,248 bytes, has MD5 \texttt{f80dde9a451e140c39b98c2e5214d93c}, and stores a float32 array of shape $16\times946\times4230$ (participants by time points by voxels). The time axis covers the first movie half supplied in this derivative: 946 TRs at 1.5 s, or 1,419 s in total. Participant numbering in the analysis follows the array order.

We fit the group-mean response and each individual response to examine the temporal organization of both levels of the data. Each participant matrix was converted to FP64, and the group response was computed by accumulating the 16 participant arrays divided by 16 in FP64. We retained the preprocessing supplied with the dataset.

All 17 fits use HP-GSBS 0.2.0 with the CPU backend, Full-Q interval scores, Prefix2D t-distance, streaming candidate evaluation, and latency mode. The scientific configuration is $K_{\max}=60$, $d_{\min}=1$, \texttt{statewise\_detection=True}, \texttt{finetune=1}, \texttt{finetune\_order=True}, and \texttt{blocksize=50}. BLAS and Numba thread counts are set to one. The host CPU is an AMD Ryzen 7 7840H under Ubuntu 22.04/WSL2 with Python 3.10.12. The cohort analysis includes one fit to the group-mean response and one fit to each of the 16 individual responses, with a total runtime of 213.10 s (Table~\ref{tab:fmrifits}).

\subsection*{Boundary matching and null construction}
For each individual, the internal boundary indices of the selected partition form a boundary sequence. To compare two participants, we find the largest possible number of one-to-one matches between their sequences, subject to an absolute difference of at most three time-point indices for each matched pair. Each match uses one boundary from each sequence, and each boundary contributes to at most one match. The sorted matching implementation was independently checked against maximum bipartite matching. If $M$ boundaries match between sequences containing $n_i$ and $n_j$ boundaries, the pairwise alignment score is $F_1=2M/(n_i+n_j)$. This normalization accounts for the different numbers of boundaries that may be detected in the two participants. The observed cohort statistic is the mean of these scores over all $\binom{16}{2}=120$ participant pairs.

To estimate the chance alignment while retaining the temporal spacing of each participant's estimated boundaries on a circular time axis, we generate a null distribution by independently shifting the boundary sequences. For each null draw, every participant receives an integer offset sampled uniformly from 1 through 945. Adding the offset to the boundary indices modulo 946 preserves the number of boundaries and their circular interboundary spacing while disrupting the original temporal correspondence between viewers. After shifting, matching uses the same absolute-distance rule on the linear recording timeline, including any boundary shifted to index zero. We then average the 120 pairwise $F_1$ scores to obtain one null cohort statistic.

Across 1,000 circular-shift permutations (seed 20261020), the null cohort-mean $F_1$ distribution has a mean of 0.3249 and a central 95\% interval of [0.3148, 0.3354]. The observed cohort-mean $F_1$ of 0.4141 exceeds the null mean by 0.0892 and is greater than every null draw, yielding a one-sided Monte Carlo $p$-value of $1/1001\approx0.001$. Each permutation shifts the 16 participants' boundary sequences independently and then averages the 120 pairwise scores, preserving their shared-participant structure in the cohort statistic.

\subsection*{Saved outputs and figure construction}
Each compact fMRI NPZ stores the selected segmentation together with the information needed to inspect alternative evaluated state counts. Table~\ref{tab:fmrifields} lists its five fields. The files \path{application_fmri/manifest.json} and \path{application_fmri/results.json} record the configuration, input identity, software version, measured times, and circular-shift scores. Table~\ref{tab:fmrifits} reports the selected state counts and runtimes for the group and all 16 participants. Seven individual fits select index 60 or 61, and the group fit selects index 60. These selections lie at the upper end of the evaluated search range. The complete saved t-distance curves support inspection of the state-count criterion and further sensitivity analyses.

\begin{table}[!ht]
\centering\small
\caption{Fields stored in each compact Sherlock segmentation file, including the selected result and the information retained across the evaluated state-count indices. Participant filenames use zero-based indices 00--15, corresponding to Participants 1--16 in the article.}\label{tab:fmrifields}
\begin{tabularx}{\textwidth}{@{}lX@{}}\toprule
Field & Content \\\midrule
\texttt{selected\_K} & Scalar index of the selected number of states \\
\texttt{boundary\_indices} & Internal boundary positions of the selected partition \\
\texttt{evaluated\_K\_indices} & State-count indices evaluated during the search, including an accepted $K_{\max}+1$ slot when present \\
\texttt{all\_bounds} & Complete boundary history indexed by state count \\
\texttt{tdists} & Complete t-distance array in the reference indexing convention; evaluated indices are listed separately \\
\bottomrule\end{tabularx}
\end{table}

For the activity panel of Fig.~6 in the main article, we select 72 evenly spaced voxel indices from the group-mean response. At each displayed time point, the activity across these voxels is spatially z-scored using the sample standard deviation, and the displayed color scale is clipped at $\pm3$. These transformations apply only to the heatmap; the segmentation itself uses all 4,230 supplied voxels. The boundary raster displays the selected boundaries of the group and individual responses, the curve panel displays all evaluated state-count indices, and the histogram displays the 1,000 null cohort means together with the observed value. The compact figure cache and associated scripts reconstruct these panels from the saved results.

\begin{longtable}{@{}lrrr@{}}
\caption{Complete Sherlock cohort fits for the arithmetic mean response and the 16 individual participants. The table gives the selected state-count index, the number of internal boundaries (one fewer than the number of states), and the measured complete-fit runtime. Seven individual selections and the group selection lie at the upper end of the evaluated search range.}\label{tab:fmrifits}\\
\toprule Matrix & Selected $K$ & Boundaries & Fit time (s) \\\midrule\endfirsthead
\toprule Matrix & Selected $K$ & Boundaries & Fit time (s) \\\midrule\endhead
\bottomrule\endfoot
Group mean & 60 & 59 & 13.5556 \\
Participant 1 & 60 & 59 & 13.0830 \\
Participant 2 & 41 & 40 & 13.1543 \\
Participant 3 & 21 & 20 & 12.4916 \\
Participant 4 & 60 & 59 & 12.3681 \\
Participant 5 & 45 & 44 & 12.4939 \\
Participant 6 & 38 & 37 & 12.2782 \\
Participant 7 & 61 & 60 & 12.3665 \\
Participant 8 & 60 & 59 & 12.3386 \\
Participant 9 & 60 & 59 & 12.2270 \\
Participant 10 & 55 & 54 & 12.3911 \\
Participant 11 & 19 & 18 & 12.9135 \\
Participant 12 & 61 & 60 & 12.6413 \\
Participant 13 & 58 & 57 & 12.4916 \\
Participant 14 & 34 & 33 & 12.3703 \\
Participant 15 & 39 & 38 & 11.8765 \\
Participant 16 & 61 & 60 & 12.0616 \\

\end{longtable}

\subsection*{Reproducing the application}
The archived results can first be checked directly, including recomputation of the null distribution from the saved boundaries. From the public repository root, the following commands verify the 17 segmentations, the observed alignment statistic, and the 1,000 null scores, and regenerate Fig.~6 from the archived display data:
\begin{lstlisting}[language=bash]
python scripts/verify_fmri.py --recompute-null
python figure_sources/figure6/draw_figure6.py
\end{lstlisting}

To perform a new fit using the versioned source dataset, first retrieve the input using the download helper, which checks the file size, MD5, array shape, and data type:
\begin{lstlisting}[language=bash]
python scripts/download_sherlock.py \
    --output external/Sherlock_AG_movie.npy
\end{lstlisting}
Install the version 0.2.0 wheel provided by the software release \citep{hpgsbs2026}. The filename is \path{hpgsbs-0.2.0-py3-none-any.whl}, and its complete-file SHA-256 is
\begin{center}\small\ttfamily
43f1819b7d8c31adea736345210550b0a5aae4a247e77e363c6650b1b5db749b
\end{center}
The public repository provides the installation instructions and \path{requirements-rerun.txt}.
With numerical-library thread counts set to one, the archived fitting script generates a new group segmentation, all 16 individual segmentations, the cohort alignment statistic, and the circular-shift scores:
\begin{lstlisting}[language=bash]
OMP_NUM_THREADS=1 OPENBLAS_NUM_THREADS=1 \
MKL_NUM_THREADS=1 NUMBA_NUM_THREADS=1 \
python scripts/run_sherlock.py \
    --input external/Sherlock_AG_movie.npy \
    --output new_sherlock_results --kmax 60 \
    --subjects all --perm 1000 --tolerance 3 \
    --null-seed 20261020
\end{lstlisting}
The new output directory records its own measured fit times and configuration.

\bibliographystyle{sn-apacite}
\bibliography{references}